# Sleeping Beauties in Medical Research
*Technological Relevance, High Scientific Impact*

**Anthony F.J. van Raan, Jos J. Winnink**
Centre for Science and Technology Studies, Leiden University, Kolffpad 1, P.O. Box 905, 2300 AX Leiden, The Netherlands.
E-mail: vanraan@cwts.leidenuniv.nl; winninkjj@cwts.leidenuniv.nl.

## *Abstract*

We investigate Sleeping Beauties (SBs) in medical research with a special focus on SBs cited in patents. We find that the increasing trend of the relative number of SBs comes to an end around 1998. However, still a constant fraction of publications becomes an SB. Many SBs become highly cited publications, they even belong to the top-10 to 20% most cited publications in their field. We measured the scaling of the number of SBs with sleeping period length, during-sleep citation-intensity, and with awake citation-intensity. We determined the Grand Sleeping Beauty Equation which shows that the probability of awakening after a deep sleep is becoming rapidly smaller for longer sleeping periods and that the probability for higher awakening intensities decreases extremely rapidly. Scaling exponents show a time-dependent behavior which suggests a decreasing occurrence of SBs with longer sleeping periods. We demonstrate that the fraction of SBs cited by patents before awakening is exponentially increasing. This finding shows that the technological time lag is becoming shorter than the sleeping time. Inventor-author self-citations may result in shorter technological time lags, but this effect is small. Finally, we discuss characteristics of an SBs that became one of the highest cited medical papers ever.

## Introduction

In science, a Sleeping Beauty is a publication that goes largely or completely unnoticed ('sleeps') for a long time and then, almost suddenly, attracts a lot of attention ('is awakened by a prince'). Garfield focused the attention on the phenomenon of 'delayed recognition' (Garfield 1970, 1980, 1989, 1990) which was linked to 'premature discovery' or 'being ahead of time' (Stent 1972). We refer to our earlier paper (van Raan 2015) for a comprehensive overview of the literature on Sleeping Beauties (SBs). In that paper we discussed the results of an extensive analysis of Sleeping Beauties in physics, chemistry, and engineering & computer science in order to find out the extent to which Sleeping Beauties are application-oriented and thus are potential Sleeping Innovations. We found that more than half of the SBs are application-oriented.

In follow-up papers (van Raan 2017a, van Raan & Winnink 2018) we took a further step by investigating whether the Sleeping Beauties in physics, chemistry, and engineering & computer science are also cited in patents, i.e., SBs that appear as scientific non-patent references (SNPR) in patents (van Raan 2017b). One of our central topics was the time lag between the publication year of the SB-SNPRs and their first citation in a patent. We found evidences that this time lag was becoming shorter in recent years. In this paper we investigate this further for medical research in order to find out how this phenomenon differs between natural science research and medical research and whether also in medical research the technological prince comes earlier than the scientific prince.

The structure of this paper is as follows. We first discuss the selection of specific sets of medical SBs, the data collection, and basic numbers and trends of identified SBs as a function of time. Next we discuss the scaling of the number of SBs with sleeping period, sleeping intensity and awake citation intensity. We then turn to the matching of the SBs with patent citation data in order to identify the SBs cited in patents (SB-SNPRs) and to analyze the time lag between publication year of the SB-SNPR and the year of the first patent citation. In this context we also focus on the technological impact of the patents that cite the medical SB-SNPRs and on the role of inventor-author combinations. In a following section we discuss the characteristics of an extremely highly cited SB-SNPR. Finally, we present a first exploration of co-citation and bibliographic coupling relations between SB-SNPRs and their citing patents.

## Identification of Sleeping Beauties

### Measurement Variables and Choice of Sets

In the foregoing papers (van Raan 2015, 2017a; van Raan & Winnink 2018) we discussed our fast and efficient search algorithm written in SQL which can be applied to the CWTS enhanced Web of Science (WoS) database with starting year 1980. With this algorithm we can tune the following four main variables: (1) *length of the sleep* in years after publication (sleeping period $s$); (2) *depth of sleep* in terms of the citation rate during the sleeping period ($c_s$); (3) *awake* period in years after the sleeping period ($a$); and (4) *awake citation-intensity* in terms of the citation rate during the awake period ($c_a$). We define $c_s$=0 as a coma, $c_s$ between 0.1 and 0.5 as a very deep sleep, and $c_s$ between 0.6 and 1.0 as a deep sleep. In the algorithm we can apply a threshold value $c_s$(max) for the citation rate during the sleeping period. For instance, if we take $cs$(max)=1.0 we cover the range from $cs$=0 (coma) to $cs$=1.0 (deep sleep). In this study we use a five-years awake period immediately after the sleeping period, i.e., $a$(max)=$a$(min)=5. Furthermore, we require that the SBs must have an awake intensity of at least, on average, 5 citations per year, i.e., $ca$(min)=5.0. For a proper analysis of the SBs, we need a total measuring period equal to sleeping period plus awakening period of five years ($a$(max)=$a$(min)=5). Clearly, the longer $s$, the less publication years we have for our investigation. For instance, if $s$=20, we need a time period 20+5=25 years, and given that 2016 is the last year of the citation measurements only SBs (with $s$=20) published between 1980 and 1992 can be taken into account. On the other hand for SBs with $s$=5 we need 5+5=10 years and thus 2007 is in this case the last publication year of the measurement.

The above definition of sleeping and awakening period can be written as follows. Given that $t_1$ is the year of publication and $c(t_i)$ is the number of citations (excluding self-citations) in any year $t_i$ then if

$$\{c(t_1) + \ldots c(t_n)\}/n \leq 1.0 \text{ and } \{c(t_{n+1}) + \ldots c(t_{n+5})\}/5 \geq 5.0$$

the sleeping time is n years in time period $[t_1, t_n]$ and the subsequent time period $[t_{n+1}, t_{n+5}]$ is the awakening period.

We identify with help of our SQL search algorithm all publications covered by the WoS in all medical research fields (see Table S1, Supplementary Material) that meet the parameters in Table 1. In this way we find all SBs with different sleeping periods, citation rates during sleep up to 1.0 and citation rates from 5.0 during the awake period of 5 years in the given range of publication years. This enables us to determine the annual numbers of these SBs. In order to identify these SBs, the search algorithm had to calculate for about 7.2 million publications their complete citation history (1980-2017) covering 230 million citations.

TABLE 1. Variables and numbers

| $s$ | $cs$(max) | $a$(max)=$a$(min) | $ca$(min) | pub years | N |
|---|---|---|---|---|---|
| 5 | 1.0 | 5 | 5.0 | 1980-2007 | 5,247 |
| 10 | 1.0 | 5 | 5.0 | 1980-2002 | 614 |
| 15 | 1.0 | 5 | 5.0 | 1980-1997 | 199 |
| 20 | 1.0 | 5 | 5.0 | 1980-1992 | 110 |
| 25 | 1.0 | 5 | 5.0 | 1980-1987 | 62 |
| 30 | 1.0 | 5 | 5.0 | 1980-1982 | 19 |

In order to give a first overall impression of quantities, we also give in Table 1 the total number of identified SBs with a specific sleeping time within the relevant range of publication years. An interesting question is: how many publications are there with the *same* sleeping characteristics as the SBs but with awake citation-intensities ($ca$) *below* the threshold used for the SBs? Thus, we identified all publications in the period 1980-2007 with $cs$(max)=1.0 during the first 5 years after publication ($s$=5) *and* $ca$(max)=4.9 during the sixth to tenth year after publication ($a$(max)=$a$(min)=5). This number is 4,281,407. Thus, the probability that a publication with no or only a few citations in the first five years after publication will become a Sleeping Beauty is about 1 thousandth.

## Basic Numbers and Trends

In this section we focus on the quantitative characteristics of SBs in more detail. The total number of publications covered by the WoS increases considerably over the entire measuring period. Obviously, if more papers in a given field are published, the number of SBs will, in principle, also increase. If however the number of SBs would increase less than the total number of publications in the given fields, then this could be an indication that the probability for a publication to become a SB decreases.

We first counted the annual number of all medical publications for the period 1980[1]-2008 The results are given in Table 2. We use 2000 as index year to calculate the growth factor. In Fig.1 we show this annual trend (again 2000 as index year). Clearly, there is an exponential growth of the medical research literature covered by the WoS of about 3% per year.

TABLE 2. Number of WoS-covered publications in the medical research fields

| publication year | number | growth factor | publication year | number | growth factor |
|---|---|---|---|---|---|
| 1980 | 149,200 | 0.51 | 1995 | 236,552 | 0.82 |
| 1981 | 155,728 | 0.54 | 1996 | 275,168 | 0.95 |
| 1982 | 164,595 | 0.57 | 1997 | 280,090 | 0.97 |
| 1983 | 171,317 | 0.59 | 1998 | 286,664 | 0.99 |
| 1984 | 176,770 | 0.61 | 1999 | 290,279 | 1.00 |
| 1985 | 184,079 | 0.63 | 2000 | 289,966 | 1.00 |
| 1986 | 191,139 | 0.66 | 2001 | 286,523 | 0.99 |
| 1987 | 195,131 | 0.67 | 2002 | 289,944 | 1.00 |
| 1988 | 201,620 | 0.70 | 2003 | 298,903 | 1.03 |
| 1989 | 209,850 | 0.72 | 2004 | 309,131 | 1.07 |
| 1990 | 214,170 | 0.74 | 2005 | 333,650 | 1.15 |
| 1991 | 220,457 | 0.76 | 2006 | 352,924 | 1.22 |
| 1992 | 224,072 | 0.77 | 2007 | 379,239 | 1.31 |
| 1993 | 224,123 | 0.77 | 2008 | 406,619 | 1.40 |
| 1994 | 224,721 | 0.77 | | | |

---

[1] The CWTS in-house version of the WoS contains publications from 1980 onwards.

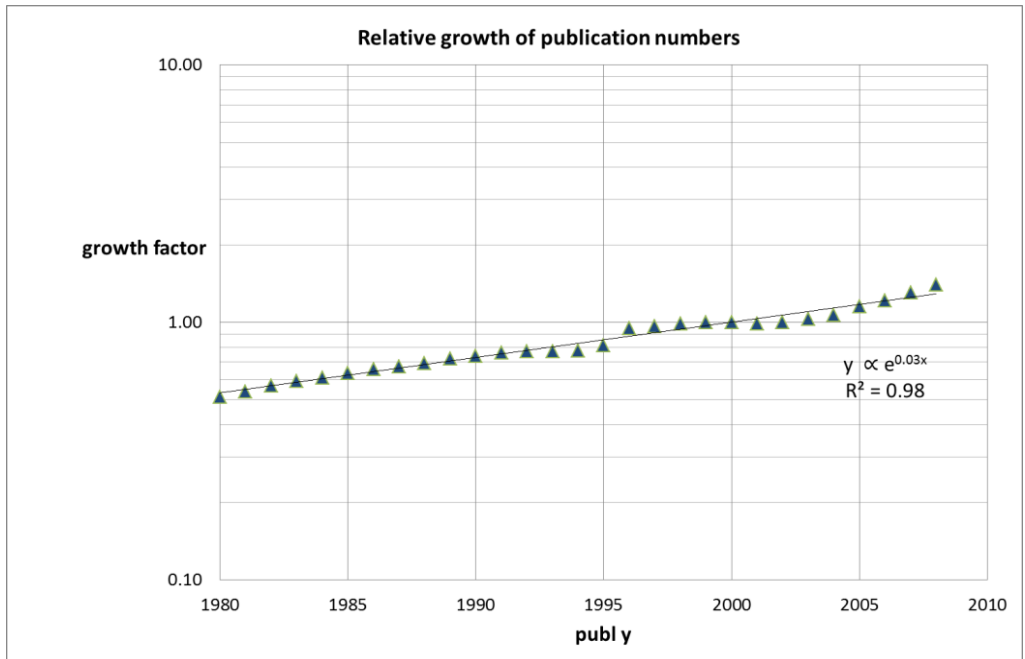

FIG. 1. Trend of the number of WoS-covered publications in the medical research fields (index: 2000).

The annual numbers of SBs are determined with our SQL search algorithm. These numbers are given in Table S2 (Supplementary Material) for all SBs with **s**=5, 10, 15, 20, 25, and 30 and for five values of **cs**(max). These numbers are what they are but if we want to find out whether the number of SBs is increasing or not, we have to *normalize* the numbers relative to the total number of all medical papers, i.e., normalization on the basis of the growth factors given in Table 2. As an example we show in Table S3 the results for SBs with **s**=5. The effect of normalization is shown in Figs. 2 and 3. In Fig. 2 the trend of the real (not-normalized) numbers of all medical publications and of the SBs with sleeping time **s**=5 is given. Notice that the number of SBs with **s**=5 is about three orders of magnitudes lower than the total number of medical publications. We see that from the late 1990's the real number of SBs does not increase anymore whereas the total number of medical papers covered by the WoS still increases. This effect is even more clear if we normalize the numbers as discussed above, see Fig. 3 where we also included the normalized numbers for SBs with **s**=10 and 15.

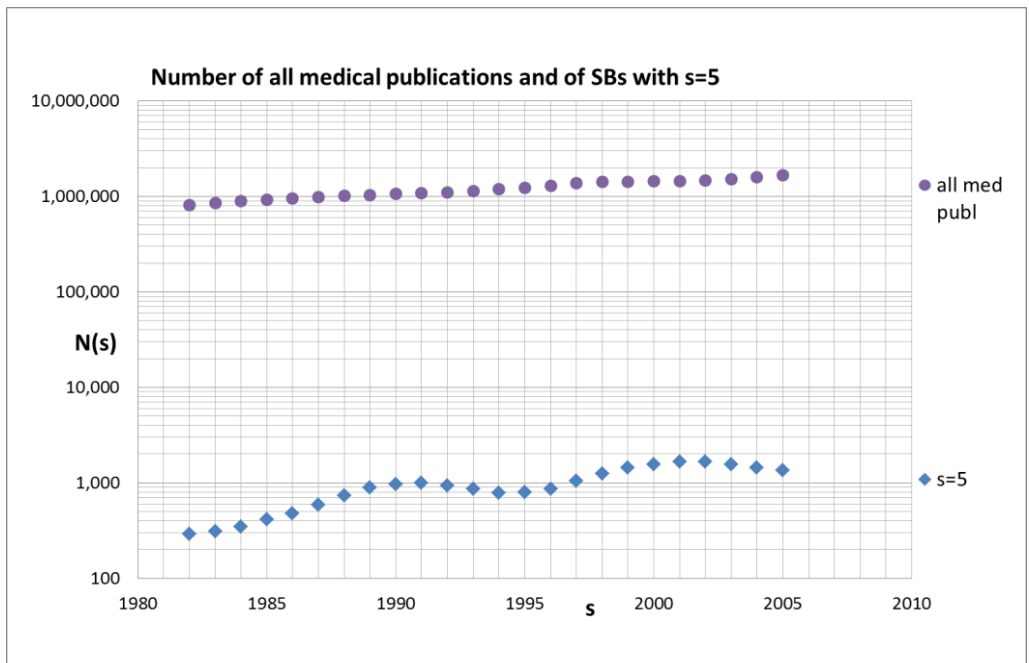

FIG. 2. Trend of the number (real, i.e. not-normalized) of all medical publications and of SBs with sleeping time $s$=5; numbers are the totals of successive five-years blocks, each five-years block is located in the figure by its middle year.

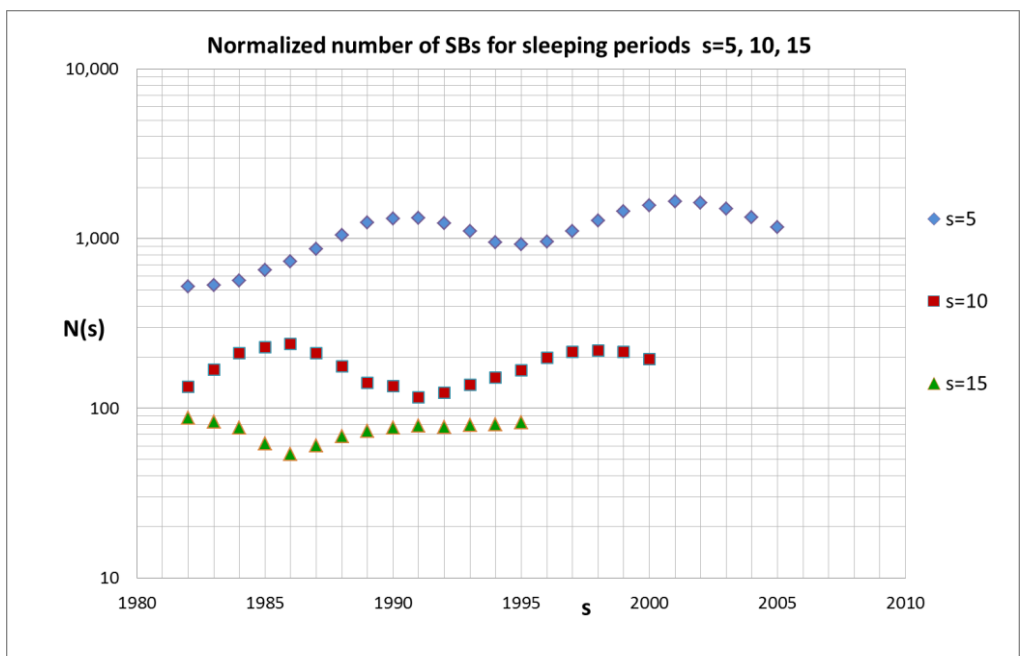

FIG. 3. Trend of the normalized numbers number of SBs in the medical research fields; numbers are the totals of successive five-years blocks, each five-years block is located in the figure by its middle year.

The SBs with $s$=5 have the shortest sleeping period and thus they can be analyzed for the most recent times, until 2007. The numbers of these SBs show considerable fluctuations during the measuring period. Looking at the normalized numbers, we find that the relative occurrence of SBs with $s$=5 doubled since the early 1980's with an average increase of 4% up to about the late 1990's. But in the more recent years the increase of the relative occurrence of SBs with $s$=5 comes to a halt. In fact, the relative occurrence of SBs with *s*=5 in 2007 is about the same as twenty years before. It confirms our observations in the natural science research fields and supports our conjecture that the expanding worldwide facilities to access scientific publications seems to have stopped increasing trends in the occurrence of SBs. However, it does not prevent that a more or less constant fraction of publications still becomes an SB.

We investigate in this study SBs with *s*=5, 10, 15, 20, 25, and 30 with a focus on the SBs with *s*=5 and 10. But obviously there are more SBs: those with sleeping times *s*=6, 7, 8, 9,….and so on. In Table S4 we present the numbers of all publications from *s*=1 to *s*=20. From *s*=5 we speak of Sleeping Beauties, but this is of course a matter of convention. Furthermore it is important to realize that there will be an overlap between the SBs found with, for instance, *s*=5 and those with *s*=6. An example shows this. Say a paper has in the twelve years starting with the publication year the following series of citation numbers: 0, 0, 0, 0, 1, 1, 5, 9, 10, 15, 11, 10. This paper will be identified as an SB with *s*=5, *cs*(max)=1.0, and *ca*(min)=5.0 because in the first 5 years after publication *cs* is (0+0+0+0+1)/5=0.2, and in the following 5 awake years *ca* is (1+5+9+10+15)/5=8.0. But this paper will *also* be identified as an SB with *s*=6, *cs*(max)=1.0, and *ca*(min)=5.0 because in the first 6 years after publication *cs* is (0+0+0+0+1+1)/6=0.33, and in the following 5 awake years *ca* is (5+9+10+15+11)/5=10.0. We leave it to the reader that this paper will also be identified as an SB with *s*=7.

Therefore the overlap between the SBs with 'multiple' sleeping times has to be determined. We will investigate this in follow-up work more precisely by adding to our search algorithm lower and upper thresholds for the sleeping time, *s*(min) and *s*(max). All in all, we find that the number of unique SBs with sleeping times *s*=5, 6, 7, 8, …….18, 19, 20 is around 10,000, thus approximately twice the number of only SBs with *s*=5.

## Sleeping Beauties Appear To Be High Impact Publications

An interesting exercise is to analyze the extent to which SBs are cited on the longer term as compared to an average publication. To that end, we determined for all medical SBs published in 2000 the annual number of citations for the period 2000-2016. Their average citation trend is presented in Fig.4 for *s*=5 with a very deep sleep (*cs*(max)=0.5, n=38), and in Fig.S1 for *s*=5 with a deep sleep (*cs*(max)=1.0, n=343) and *s*=10 also with a deep sleep (*cs*(max)=1.0, n=43) together with the citation trend of an average medical publication. We determined for each SB the total number of citations from 2000 up till now (September 23, 2018). Next we analyzed where these total numbers of citations are located in the citation distribution of the entire set of all medical research papers published in 2000. We find that of the SBs with *s*=5 and *cs*(max)=0.5, 55% belongs to the top-10% and all to the top-25% of the citation distribution of all medical research papers. For *s*=5 and *cs*(max)=1.0 we find practically the same results: 57% belongs to the top-10% and almost all (99%) to the top-25%. Of the SBs with *s*=10 12% belongs to the top-10% and all belong to the top-25%. Therefore perhaps an unexpected finding is that, in general, Sleeping Beauties appear to be highly cited publications in the longer term. But we also find another interesting phenomenon: the average SB citation trends shows that the awakening process appears to be characterized by a sleep that becomes lighter and lighter, i.e., a slow and small increase of the citation intensity during the sleep period, followed by an abruptly rapid increase of the citation impact.

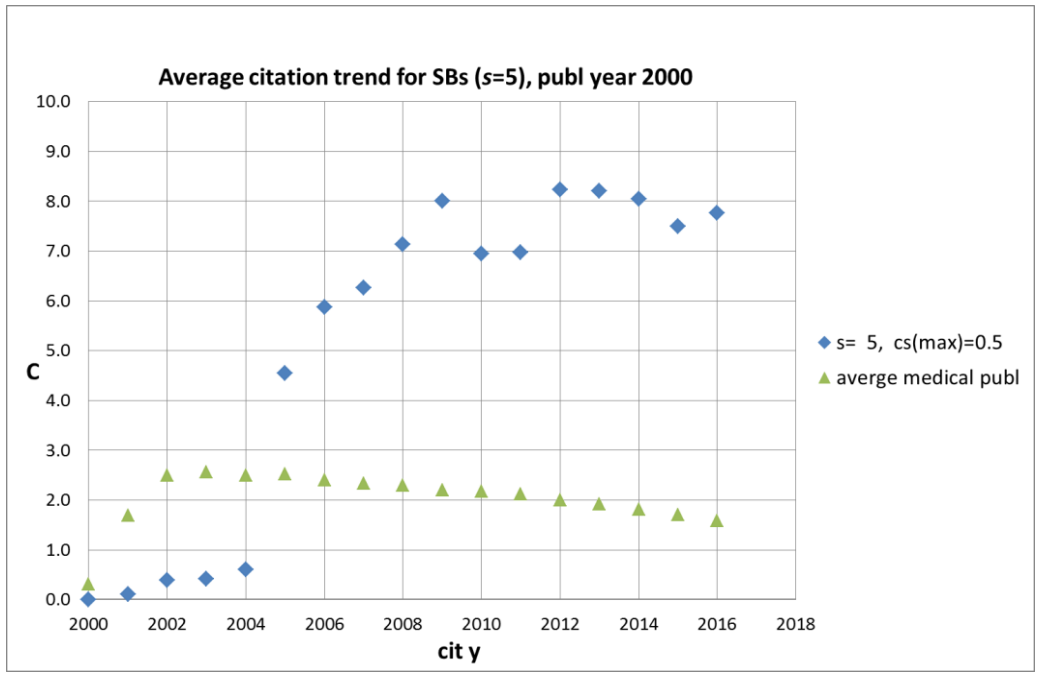

FIG. 4.  Average citation trend in the period 2000-2016 for SBs with *s*=5 and *cs*(max)=0.5 as well an average medical publication with publication year 2000.

## Scaling Phenomena and Grand Sleeping Beauty Equation

In the foregoing section we presented an overall picture of numbers and trends. We will now go into more detail by investigating the dependence of the number of SBs on the three main variables: sleeping period (*s*), during-sleep citation-intensity (*cs*), and awake citation-intensity (*ca*). In this way, we are able to determine the probability that an SB occurs with specific values of the three above variables occurs. In this sense we go back to our first SB measurements (van Raan 2004) where we used the empirical data to construct a 'Grand Sleeping Beauty Equation' (GSBE).

### Scaling with Sleeping Period

First we determine the *number of SBs as a function of sleeping time (**s**)*. As explained earlier, we have to be careful: for SBs with *s*=5 we have publication years in the long measuring period 1980-2007 available for counting. But, for instance, for *s*=15 we have the measuring period 1980-1997 and for our extreme case, SBs with *s*=30 only 1980-1982 is available as measuring period. Thus, in order to find reliable probabilities of occurrence, we have to take measuring periods which cover the same time period of publication years for SBs with different sleeping times. Moreover, it is very well possible that these probabilities are different for the early 1980's as compared to recent years. In order to investigate this in detail, we apply the following procedure. In a first step we measure the numbers of SBs for the different sleeping periods for all possible publication years (we use successive 5-year blocks of the publication years in order to have a sufficiently large numbers of SBs). The results are shown in Table 3. Because we will compare SBs with *different* sleeping periods in the *same* measuring periods we can use real (not-normalized) numbers.

TABLE 3. Number of SBs of different sleeping times and for the given range of publication years. In all cases $cs$(max)=1.0.

|    | 1980-1984 | 1981-1985 | 1982-1986 | 1983-1987 | 1984-1988 | 1985-1989 | 1986-1990 | 1987-1991 | 1988-1992 | 1989-1993 | 1990-1994 | 1991-1995 |
|----|-----------|-----------|-----------|-----------|-----------|-----------|-----------|-----------|-----------|-----------|-----------|-----------|
| 5  | 291 | 312 | 344 | 410 | 477 | 591 | 731 | 896 | 970 | 989 | 936 | 863 |
| 10 | 75  | 100 | 129 | 144 | 155 | 143 | 123 | 102 | 100 | 87  | 94  | 107 |
| 15 | 49  | 49  | 47  | 39  | 35  | 41  | 48  | 53  | 57  | 59  | 59  | 62  |
| 20 | 26  | 34  | 36  | 43  | 48  | 52  | 50  | 56  | 56  |     |     |     |
| 25 | 35  | 34  | 43  | 41  |     |     |     |     |     |     |     |     |

|    | 1992-1996 | 1993-1997 | 1994-1998 | 1995-1999 | 1996-2000 | 1997-2001 | 1998-2002 | 1999-2003 | 2000-2004 | 2001-2005 | 2002-2006 | 2003-2007 |
|----|-----------|-----------|-----------|-----------|-----------|-----------|-----------|-----------|-----------|-----------|-----------|-----------|
| 5  | 779 | 794 | 866 | 1,043 | 1,246 | 1,428 | 1,568 | 1,660 | 1,662 | 1,568 | 1,449 | 1,342 |
| 10 | 124 | 144 | 178 | 202   | 214   | 212   | 194   |       |       |       |       |       |
| 15 | 66  | 71  |     |       |       |       |       |       |       |       |       |       |
| 20 |     |     |     |       |       |       |       |       |       |       |       |       |
| 25 |     |     |     |       |       |       |       |       |       |       |       |       |

Next, we examine the relations between the number of SBs for different sleeping periods. Given the very low numbers for the longest sleeping times (see Table 3) we make the analysis for the sleeping periods $s$=5, 10 and 15. The results are shown in Fig. 5. In order not to overload the figure we take the data for 1980-1984, 1982-1986, and so on. We find a power law for the scaling of the number of SBs with $s$, with an exponent $\alpha$ starting around -1.70 in the early 1980's and a rapid increase to around -2.50.

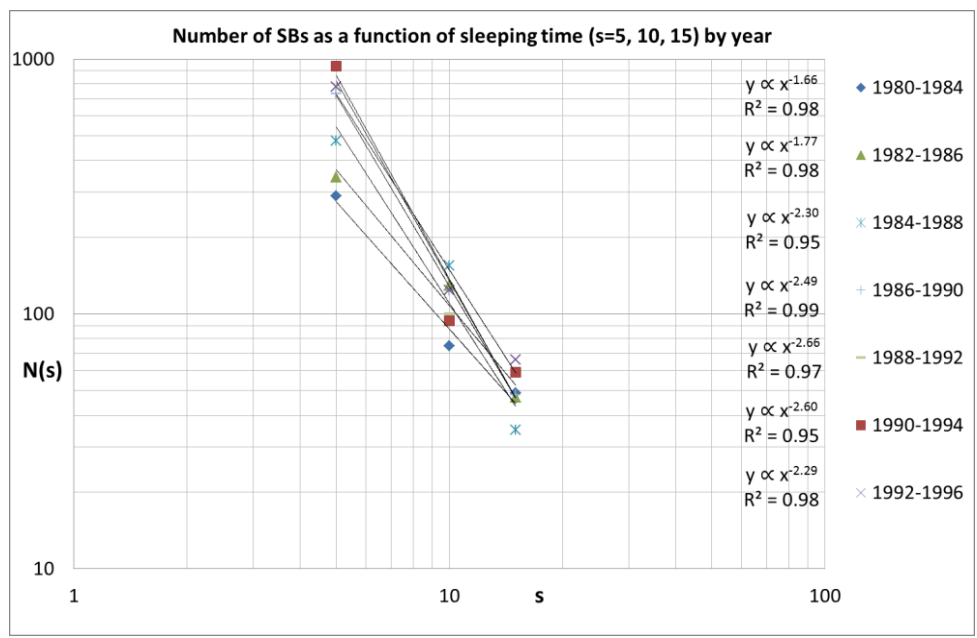

FIG. 5. Number of SBs for $s$=5, 10, 15 between 1980 and 1996.

Finally, the results for the last five publication-year blocks (1994-1998,….,1998-2002) covering $s$=5 and 10 are shown in Fig.6. Again we see a further continuation of the increasing exponent trend up $\alpha$=-3.02 (1998-2002).

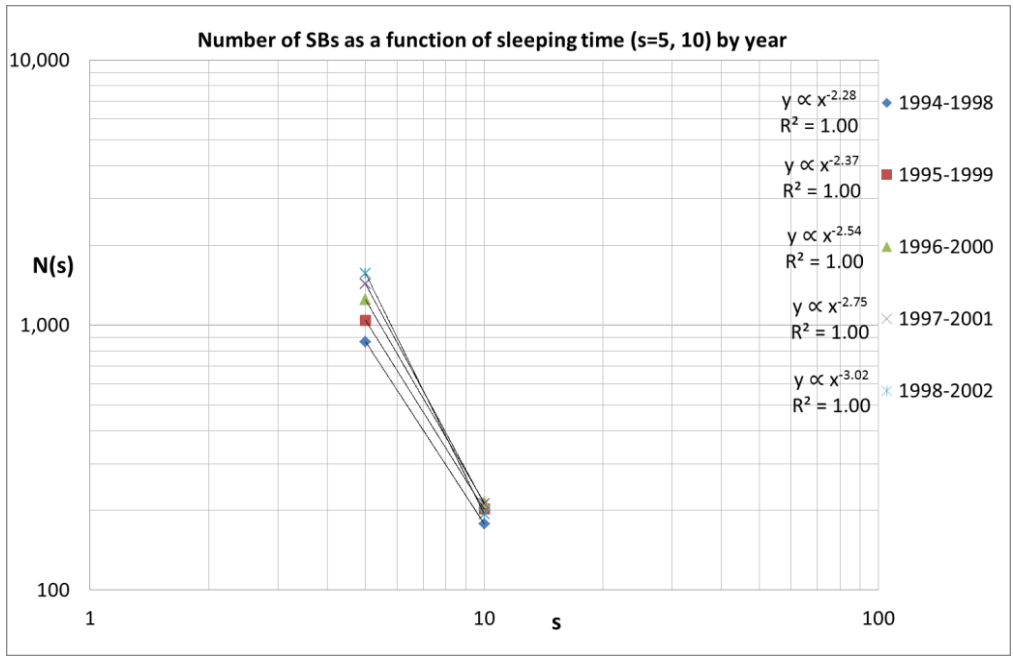

FIG. 6. Numbers of SBs for *s*=5 and 10, between 1994 and 2002.

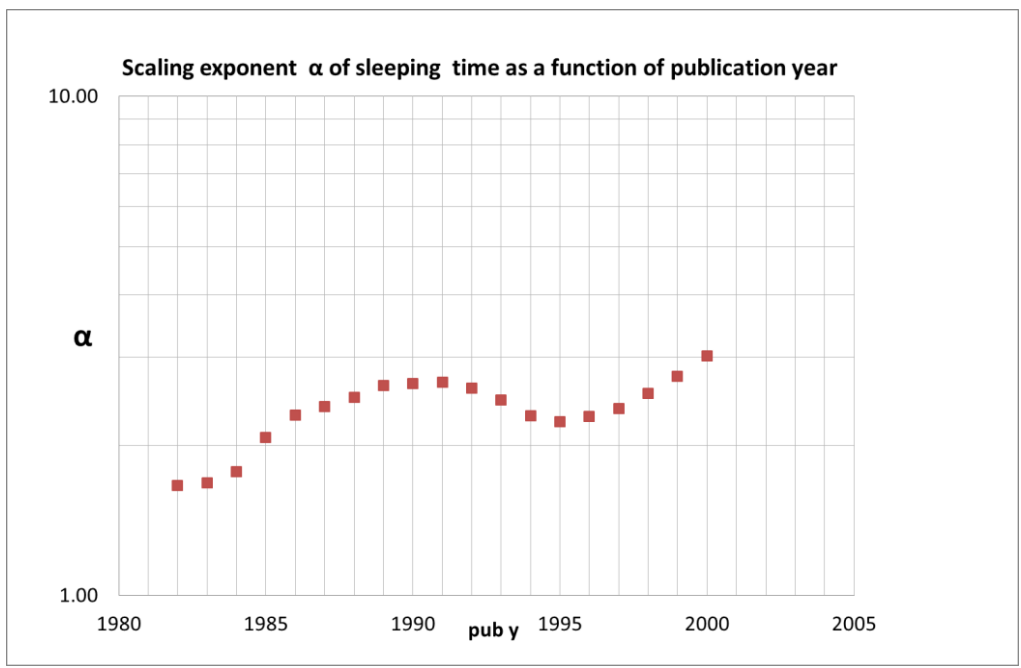

FIG. 7. Scaling exponent *α* for the whole measuring period 1980-2000 as a function of publication year; numbers are the averages of successive five-years blocks, each five-years block is located in the figure by its middle year (because of the logarithmic scale we take the absolute values of *α*).

The results are remarkable: the power law scaling of the number of SBs as a function of sleeping time (*s*=5, 10, 15) increases in the course of time, be it with considerable fluctuations. We find that in the 1980's the scaling of the numbers of SBs as a function of sleeping time has an exponent *α* of around -1.7. At the end of our measuring period however this exponent has almost doubled to around -3.0, see Fig. 7. The most probable cause of this change is that the number of SBs with a longer sleeping time decreases.

The absolute value of the number of SBs with *s*=5 influences the power law exponent *α* and therefore the fluctuations in *α* strongly resemble those of the number of SBs with *s*=5 SBs, as comparison of Fig. 3 with Fig. 7 shows. For the last publication-year block the number of SBs scales approximately as

$$N(s) = A \cdot s^{-3.0} \tag{1}$$

The coefficient *A* is also determined by the empirical results; for instance, for the period 1998-2002 *A*=200,724. It is interesting to compare these measurements with those in our first Sleeping Beauty paper (van Raan 2004). This comparison has its restrictions because the 2004-study concerned all disciplines. We take the same range of publication years as in the 2004-study (1980-1997) and determine for our medical SBs the power law exponent for the relation between the number of SBs with *s*=5 and 10 with sleeping period. We find *α*=-2.5 which is in good agreement with the results of the 2004-study where we found a power law scaling with exponent -2.7. Our observations summarized by Eq.(1) imply that if a publication is *twice as longer in deep sleep*, the probability that it awakes with, on average, at least five citation per year during five years, is about an *order of magnitude less*. In other words, the longer the sleeping period, the less probable it is that a publication will awake. Indeed a finding that we can intuitively recognize.

## Scaling with Sleeping Intensity

Next we analyze the relation between the *number of SBs and sleeping intensity*, i.e., citation-intensity during sleep (*cs*). This has to be done for each sleeping period separately because SBs with a relatively short sleeping period (*s*=5) may have different citation-intensity distributions as compared to SBs with a much longer sleeping period (*s*=10 and longer). We investigated the scaling of the number of SBs with citation-intensity *cs*=0.2, 0.4, 0.6, 0.8, and 1.0. It is very well possible that citation distributions of SBs will change in the course of time, particularly in a long observation period as used in this study (1980-2017). Therefore we also analyzed these citation-intensity distributions as a function of the SB publication year. As an example we show in Fig.8 the results for the SBs with *s*=5 for the years 2001-2007. In order to avoid low numbers we again used overlapping 5-years blocks for the publication years. We see that the sleeping-intensity scaling-exponent *β* is around 1.20. In Table S5 (Supplementary Material) we present the results for all years of the measuring period 1980-2007. From these data we deduced the time-dependent development of the sleeping intensity scaling exponent *β*, see Fig.9.

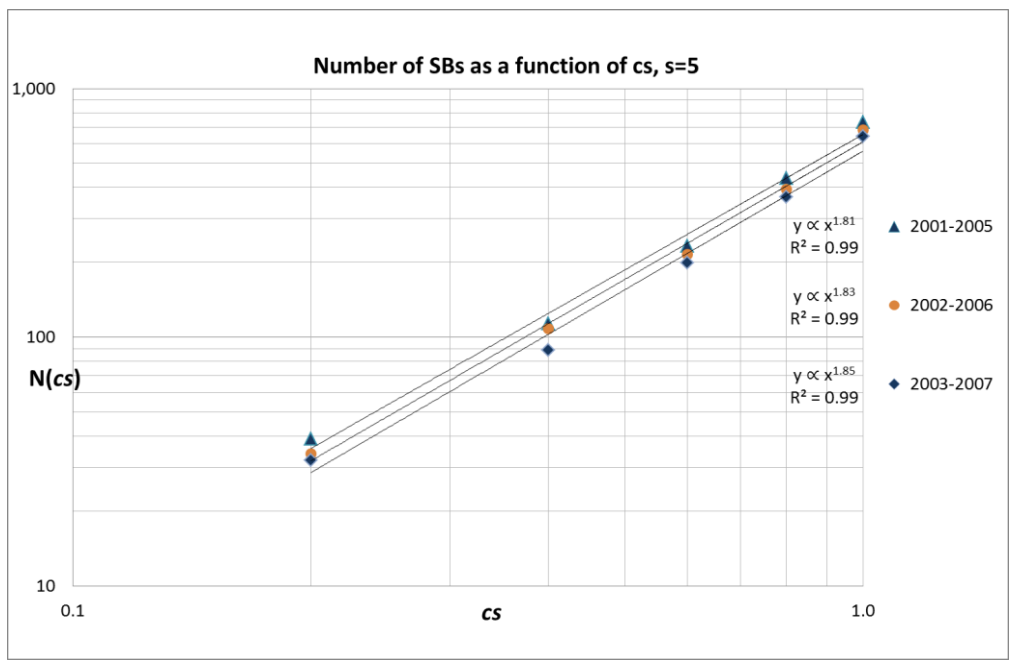

FIG. 8. Number of SBs ($s$=5) for successive during-sleep citation-intensity intervals, 2001-2007 (numbers are located with the middle values of the intervals).

Remarkably, scaling exponent $\beta$ fluctuates for a long time -in the period 1980-2000- around +1.40 ($sd$=0.08). Then it increases considerably and reaches a value around +1.80 ($sd$=0.04) as we see in Fig.9. This implies that the SBs tend to a 'less deep sleep'. Thus, in this recent period the number of SBs scales approximately as (coefficient $B$ follows from the empirical data, for the period 2003-2007 $B$=560.37):

$$N(cs) = B \cdot cs^{+1.8} \qquad (2)$$

In our 2004-study we found a during-sleep scaling exponent (for all disciplines together) of around +2.5. For SBs with $s$=10 we find that the during-sleep scaling exponent fluctuates during the whole measuring period (1980-2002) around $\beta$=1.43 ($sd$=0.18), see Table S6 and Fig.S2 (Supplementary Material).

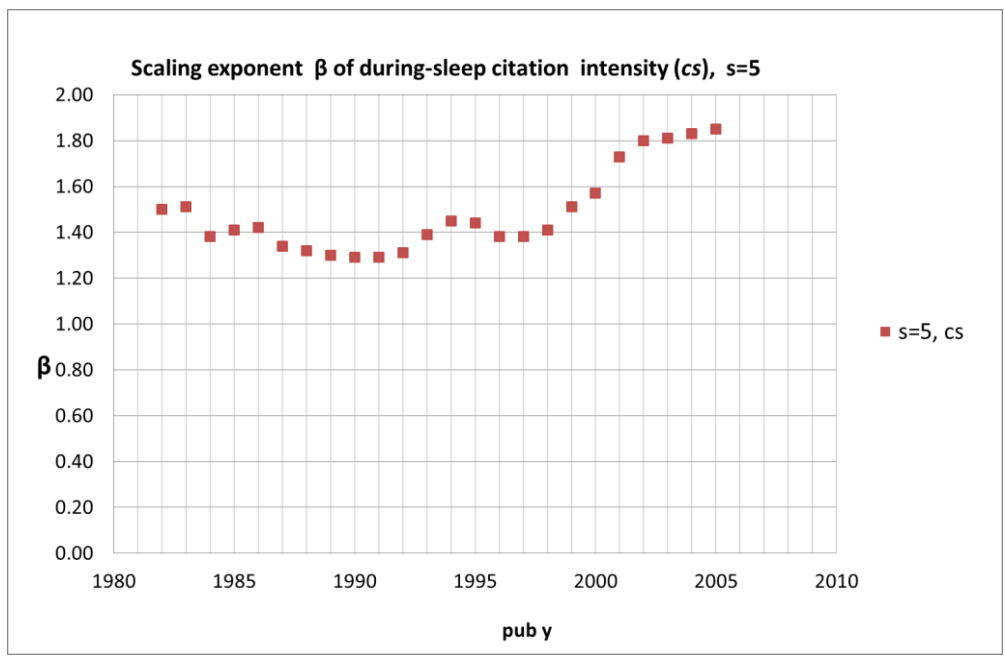

FIG. 9. Scaling exponent β of during-sleep citation intensity, SB-SNPRs with $s$=5; numbers are the averages of successive five-years blocks, each five-years block is located in the figure by its middle year.

Our observations summarized by Eq.(2) imply that the probability to find a publication that has twice as much citations in deep sleep (e.g., 4 versus 2 in five years sleep), is about factor 2.5 higher. In other words, the less deep the sleep, the more SBs will be found, because we are moving toward 'normal' publications.

## Scaling with Awake Citation-Intensity

The third step is to analyze the relation between the number of SBs and awake citation-intensity ($ca$). Again this has to done for each sleeping period separately because SBs with a different sleeping periods may have different citation-intensity distributions during the awake period. We investigated the scaling of the number of SBs with successive awake citation-intensity intervals: $5.0 \leq ca \leq 6.0$, $6.0 < ca \leq 7.0$, $7.0 < ca \leq 9.0$, $9.0 < ca \leq 11.0$, $11.0 < ca \leq 13.0$, $13.0 < ca \leq 15.0$, and $15.0 < ca \leq 17.0$. Also here we analyzed these citation intensity distributions as a function of publication year. Fig.10 shows as an example the results for the SBs with $s$=5 for the publication years 2001-2007 (divided in three overlapping five-years blocks). We notice the very steep decrease of the number of SBs as a function of $ca$, with exponent γ around -6. As far as we know this is the steepest scaling exponent found in bibliometric research. For instance, in the case of measuring period 2003-2007 the number of SBs scales approximately as:

$$N(ca) = C \cdot ca^{-6.1} \qquad (3)$$

Given the steepness of the curve the value of coefficient $C$ is large, for 2003-2007 it is around $27*10^6$. In Table S7 (Supplementary Material) we present the results for all publication years in the measuring period 1980-2007. From these data we deduced the time-dependent development of the awake citation-intensity scaling exponent $\gamma$, see Fig. 11. The scaling of the awake citation-intensity shows an increase starting with $\gamma$ around -5.0 in the 1980's but since the second half of the 1990's it fluctuates around -6.0. The most plausible explanation for this phenomenon is that in recent times there are relatively less SBs with higher awake citation-intensity. For SBs with $s$=10 (data in Table S8) we find that the trend of the awake citation-intensity exponent $\gamma$ is rather constant and fluctuates around -4.33 ($sd$=0.63), see Fig.S3. These values are lower than those in our 2004-study where we found (for all disciplines together) an exponent of -6.6.

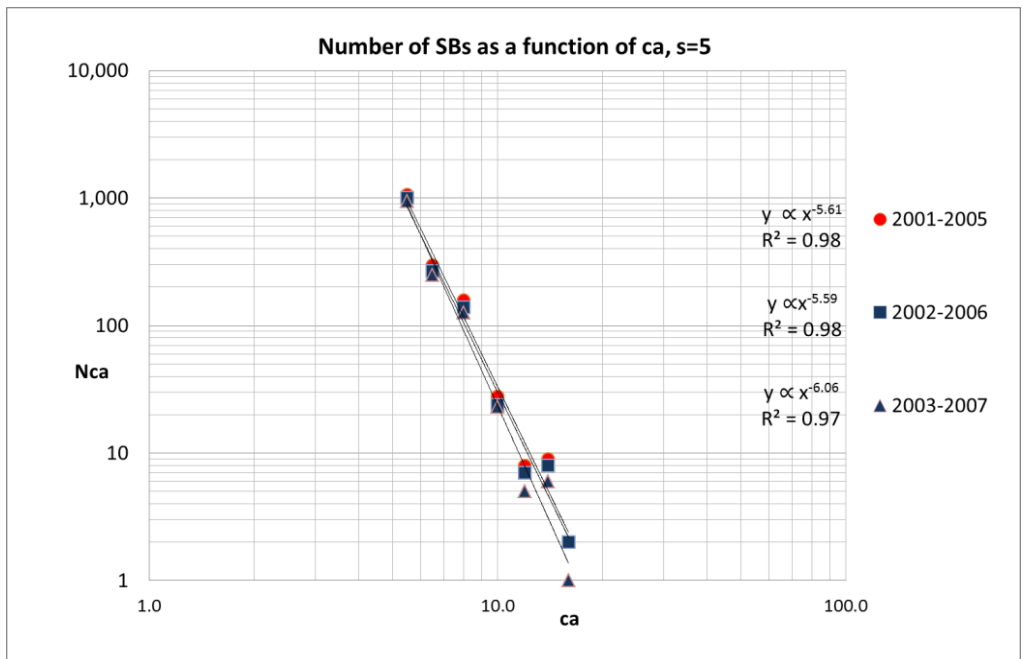

FIG. 10. Number of SBs ($s$=5) for successive awake citation-intensity intervals, 2001-2007.

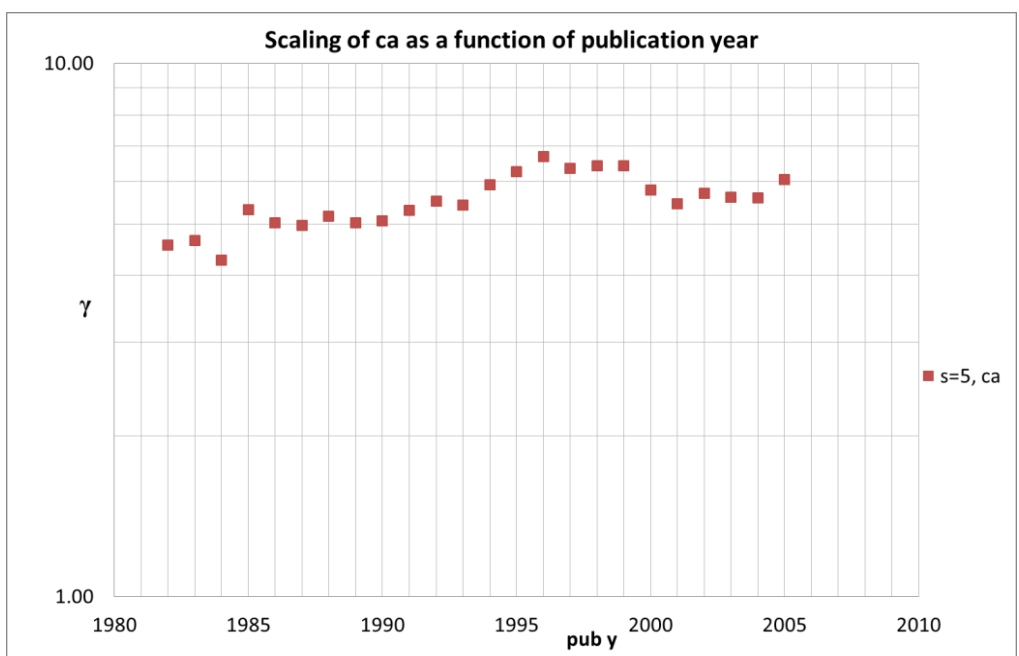

FIG. 11. Scaling exponent $\gamma$ for the whole measuring period 1980-2007 (because of the logarithmic scale we take the absolute values of $\gamma$).

All in all these observations summarized by Eq.(3) mean that the probability for a higher awake citation-intensity decreases very rapidly. For instance, the probability that a SB during the awake period will have a citation intensity twice as large as another SB (e.g., 10 versus 5 citations per year in the five years awake period), is about a factor 50 lower. Taking the empirical findings for the scaling with sleeping period length and with during-sleep citation-intensity together with the results for the scaling with the awake citation-intensity, we find for the *Grand Sleeping Beauty Equation* (see van Raan 2004) approximately:

$$N = f(s,cs,ca) \propto s^{-3.0} \cdot cs^{+1.8} \cdot ca^{-6.1} \qquad (4)$$

Eq.(4) gives (after determination of a constant factor given by the coefficients of the above discussed scaling with *s*, *cs*, and *ca*) the number of Sleeping Beauties for any sleeping time, during-sleep intensity and awake citation-intensity, and particularly the dependency on these variables. Summarizing we find the following main characteristics of Sleeping Beauties in the scientific literature. First, the strong negative exponent connected to sleeping time shows that the probability of awakening after a deep sleep is becoming rapidly smaller for longer sleeping periods. Second, the probability for higher awakening intensities decreases extremely rapidly.

## Technological Impact of Sleeping Beauties

Number and Distribution of Patent Citations

Patents are documents with a legal status to describe and claim technological inventions. Similar to scientific publications, patent documents contain references. These references are aimed at proving novelty in view of the existing technological developments. These references mainly relate to earlier patents ('prior art') and to a lesser extent to non-patent items, a major part of which are citations to scientific publications (van Raan 2017b), the 'scientific non-patent references' (SNPRs). References in scientific publications are the sole responsibility of the authors, but references in patents come from two sources: the inventors as well as the patent examiners. SNPRs generally represent an important bridge between science and technology (Narin, Hamilton and Olivastro, 1997) but they do not necessarily indicate the direct scientific basis of the invention described in the patent. Nevertheless, many studies (e.g., Callaert et al. 2014) emphasize the importance of further research of the role of SNPRs in relation to the patented technological invention. In this study we elaborate further on our previous work on SBs that are SNPRs (SB-SNPRs) with a particular focus on medical research and on as recent as possible SBs.

Patent data were collected by searching the EPO Worldwide Patent Statistical Database (PATSTAT), Spring 2016 version. In order to find out whether an SB is cited by patents, we matched all SBs on the basis of their WoS UT-codes with the citations given in patents. For more details we refer to Winnink and Tijssen (2014). We group patents describing the same invention in 'patent families'[2] to prevent double counting. But even then it is possible that an SB is matched several times within the same patent family because there may be multiple versions with different patent filing years. Therefore it is necessary to check for each SB the occurrence of multiple versions and to correct for this. In the case of multiple versions, we take the version with the first filing year. For the sake of brevity we will speak about patents instead of patent families.

---

[2] A patent family is a set of, in legal terms, equivalent patents that describe one and the same invention. This is for instance the case when the same invention is patented in more than one country. In this article we use the term 'patent' also for 'patent families'.

We created a database in which the SBs are uniquely linked with their (first filed) citing patents. This database allows us to determine how many, and which, SBs of sleeping periods (*s*=5, 10, 15, 20, 25) are cited by how many, and which, patents within a specified period *after publication*. This is important because, for example, if SBs with *s*=5 have citing patents within the first 5 years after their publication, it means that these SBs are cited by patents during their (scientific) sleeping period. Thus, in that case a 'technological prince' is earlier than, or at least at the same time as, the 'scientific prince'. A period of 10 year after publication is the longest period that is applicable to all SBs: the most recent SBs are published in 2007 and a complete 10 years period ends in 2016, the last year of our patent matching measurements. Obviously, for SBs published in 1980 the citation history can be much longer, up to 2016, which means 37 years. Thus, taking the whole available period allows us to study how the technological impact of SBs (based on citations by patents) changed over time. We remind that we analyze SBs with sleeping periods (*s*=5, 10, 15, 20, 25) in order find general characteristics. A complete analysis would involve all possible sleeping times *s*=6, 7, 8, 9,….and so on.

In Table 4 we present the number of SBs cited by patents. This table can be read as follows, we take as an example the SBs with a short (*s*=5) sleep. We see that there are in total (1980-2007) 5,261 SBs of which 663 are cited by patents over the whole period (right-most block in the table). Thus, 12.6% of all SBs with *s*=5 is an SB-SNPR. These 663 SB-SNPRS are cited in total 2,074 times, by 1,745 patents. With a period of 10 year after publication of an SB, it is found that 492 SBs (9.4% of all SBs with *s*=5) are cited 1,279 times by 1,116 patents. In the case of a period of 5 year after publication of an SB with *s*=5, all identified patent citations will be *within the sleeping period* of the SBs. We find that 313 SBs are cited within their sleeping period (6.0% of all SBs with *s*=5) 565 times by 508 patents. Given that 313 is about 47% of 663, we conclude that almost half of all SB-SNPRs (*s*=5) are cited by patents within their sleeping period. As discussed above, the data in Table 4 relate to the total measurement period 1980-2007. In the next section we will analyze the time-dependence of the citations of SBs by patents.

TABLE 4. Number of SBs cited by patents

| | SBs total | SBs cited within 5y | % | pat cit | pats | SBs cited within 10y | % | pat cits | pats | SBs cited up to 2017 | % | pat cits | pats |
|---|---|---|---|---|---|---|---|---|---|---|---|---|---|
| s= 5 | 5,247 | 313 | 6.0% | 565 | 508 | 492 | 9.4% | 1,279 | 1,116 | 663 | 12.6% | 2,074 | 1,745 |
| s=10 | 614 | 26 | 4.2% | 36 | 34 | 67 | 10.9% | 122 | 114 | 116 | 18.9% | 417 | 331 |
| s=15 | 199 | 7 | 3.5% | 8 | 7 | 9 | 4.5% | 21 | 19 | 32 | 16.1% | 79 | 74 |
| s=20 | 110 | 2 | 1.8% | 2 | 2 | 3 | 2.7% | 3 | 3 | 13 | 11.8% | 20 | 18 |
| s=25 | 62 | 0 | 0.0% | 0 | 0 | 0 | 0.0% | 0 | 0 | 6 | 9.7% | 10 | 10 |

A substantial amount of SBs is cited by more than just one patent. Some SBs are cited by more than 10 or even more than 20 patents. It is clear that these highly cited SBs are interesting cases for further analysis, and we will come back later to this issue. In Figs.12 and 13 we present the number of SB-SNPRs (Fig.12 for *s*=5 and Fig.13 for *s*=10) as a function of number of patent citations, with citation windows 5 and 10 year after publication. We see that similar to citations given by publications, also the number of citations given by patents is characterized by a skewed distribution. For our medical SB-SNPRs we find similar distributions as in the case of the natural sciences and engineering (van Raan & Winnink 2018).

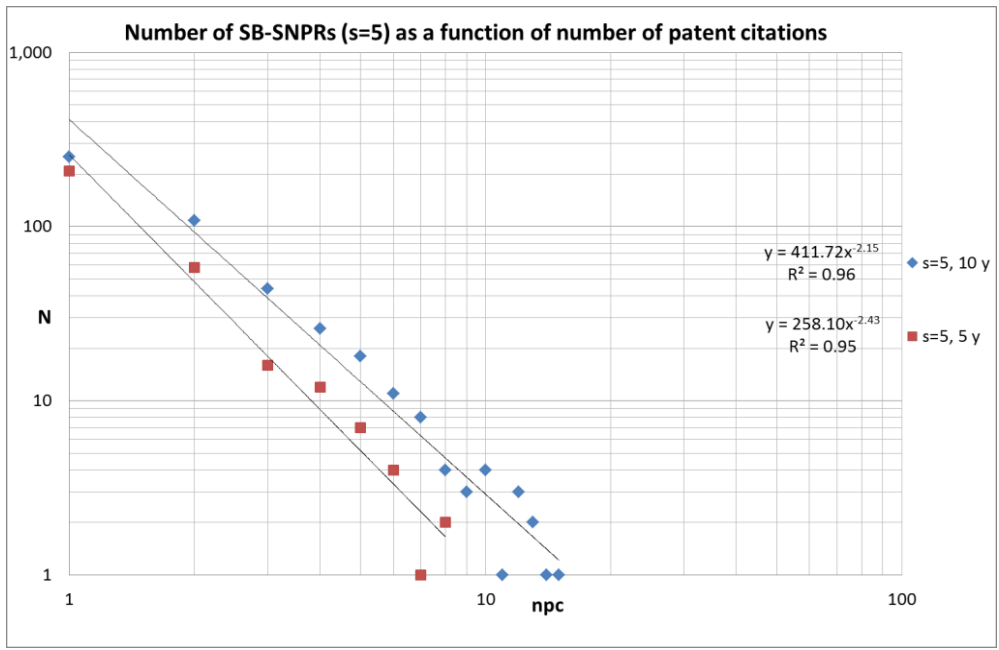

FIG. 12. Number of SBs with *s*=5 cited by patents as a function of number of patent citations, citation windows 5 and 10 year.

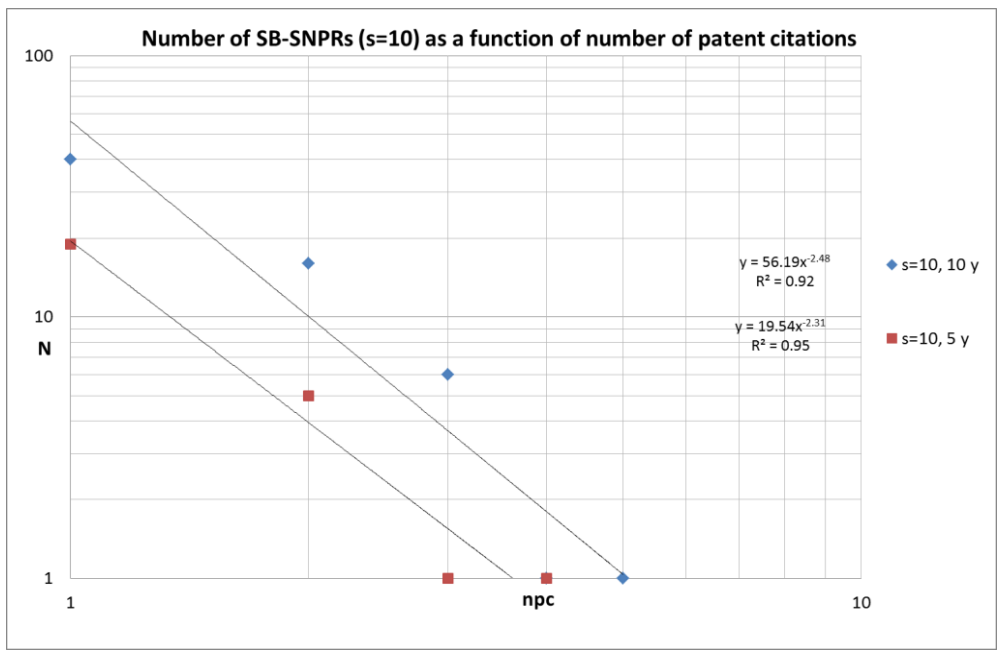

FIG. 13. Number of SBs with *s*=10 cited by patents as a function of number of patent citations, citation windows 5 and 10 year.

Overall, we find for the SB-SNPRs with the relative short sleep ($s$=5) that within a period of *10 years after publication* 26% is still not cited (thus they are cited by patents more than 10 years after their publication), about half (54%) is cited by 1 or 2 patents, and about 20% is cited by 3 of more patents. Within a period of *5 years after publication*– which means patent citations *within the sleeping period*- about half (53%) is not cited yet, 41% of the SB-SNPRS is cited by 1 or 2 patents, and 7% is cited by 3 of more patents. For the SB-SNPRs with the long sleep ($s$=10) we find that within a period of *10 years after publication* –which is in this case *within the sleeping period*- 42% is still not cited, about half (48%) of is cited by 1 or 2 patents, and about 10% is cited by 3 of more patents. Within a period of *5 years after publication*– which means for the SB-SNPRS with $s$=10 widely within the sleeping period- 78% is not cited yet, 20% of the SB-SNPRS is cited by 1 or 2 patents, and only 2% is cited by 3 of more patents.

For the still longer sleeping periods ($s$=15, 20, …) the numbers are too small for an accurate determination of the distribution. We see in Figs.12 and 13 that the number of SB-SNPRs as a function of the number of patent citations scales with a power law exponent between -2.0 and -2.5. For the SB-SNPRs with $s$=5 (Fig.12) the distribution for patent citations within 5 years after publication seems to decrease faster than the distribution for patent citations within 10 years. This suggests that it becomes increasingly unlikely to receive more patent citations within 5 years after publication. For the SB-SNPRs with $s$=10 (Fig.13) we do not see this effect, but here the numbers are already too small to make a reliable assessment of the details of the distribution.

## Scientific or technological awakening?

The distributions presented in Figs.12 and 13 relate to the total number of SBs in the entire measuring period (1980-1997). As discussed earlier, it is very well possible that properties of SBs and of SB-SNPRs in particular, change over time. This is indeed the case, and quite dramatically: Figs.14 and 15 shows the fraction of SB-SNPRs (Fig.14 for $s$=5 and Fig.15 for $s$=10) that are cited at least once, changes in the course of time. In the 1980's about 50% of the SB-SNPRs with $s$=5 and about 20% of the SB-SNPRs with $s$=10 were cited within 10 years after publication. But in the first decade of this century almost all SB-SNPRs (both $s$=5 as well as 10) are cited within 10 years. For SB-SNPRs with $s$=5 the fraction for at least one patent citation within 5 years after publication, and thus within their sleeping period, is in the 1980's about 25%. But about 20 years later, in the first decade of this century, the situation has changed drastically: 70-80% of SB-SNPRs are cited within their sleeping period. For SB-SNPRs with the longer sleep period $s$=10 the fraction for at least one patent citation within 5 years after publication is in the 1980's about 10% and in the first decade of this century it is 40%, again a substantial change. We also notice in Figs.14 ($s$=5) and 15 ($s$=10) that the fraction of SBs cited by patents within 5 years increases more rapidly than within 10 years.

All these observations point in one and the same direction: in a rapidly increasing pace publications that still 'sleep' scientifically do already have a technological impact within their sleeping period. We must however be cautious about concluding that a technological awakening is more and more likely to occur than a scientific one. The average SB citation trends (Fig.4) show that the awakening process appears to be characterized by a sleep that becomes lighter and lighter, i.e., a slow and small increase of the citation intensity during the sleep period (but still meeting the requirement $cs$(max)=1.0), followed by an abruptly rapid increase of the citation impact. What we can conclude is that technological impact of SBs in the form of citations by patents will take place in an increasingly shorter period of time after publication, and more and more in the 'formal' sleep period of an SB.

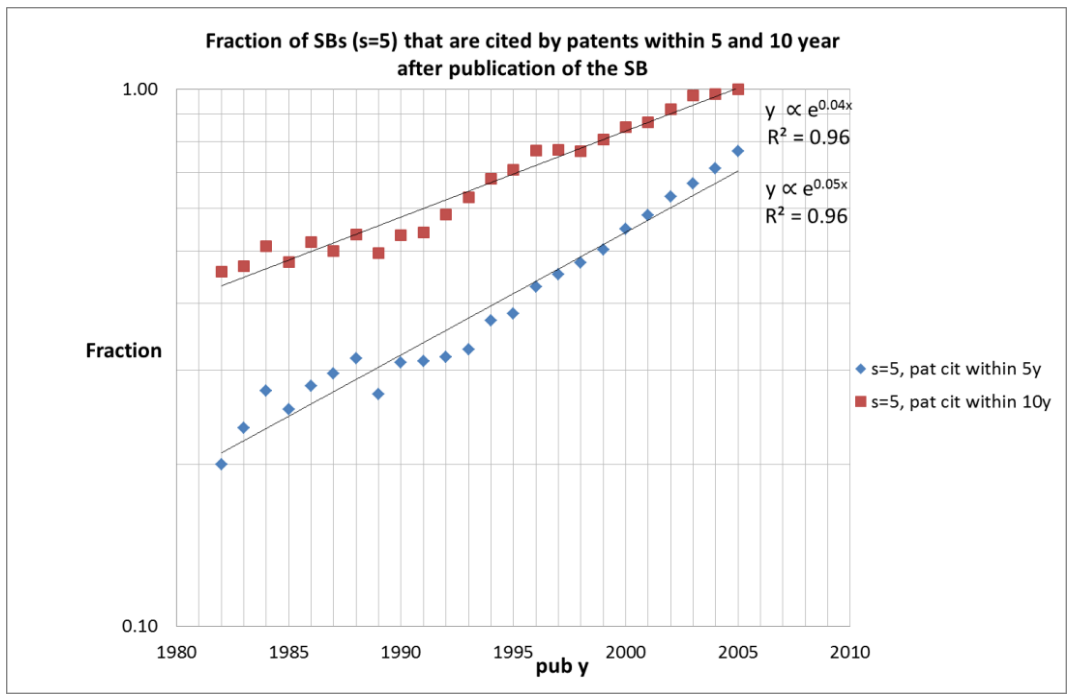

FIG. 14. Fraction of SBs with *s*=5 that are cited by patents within 5 and 10 year after publication of the SB.

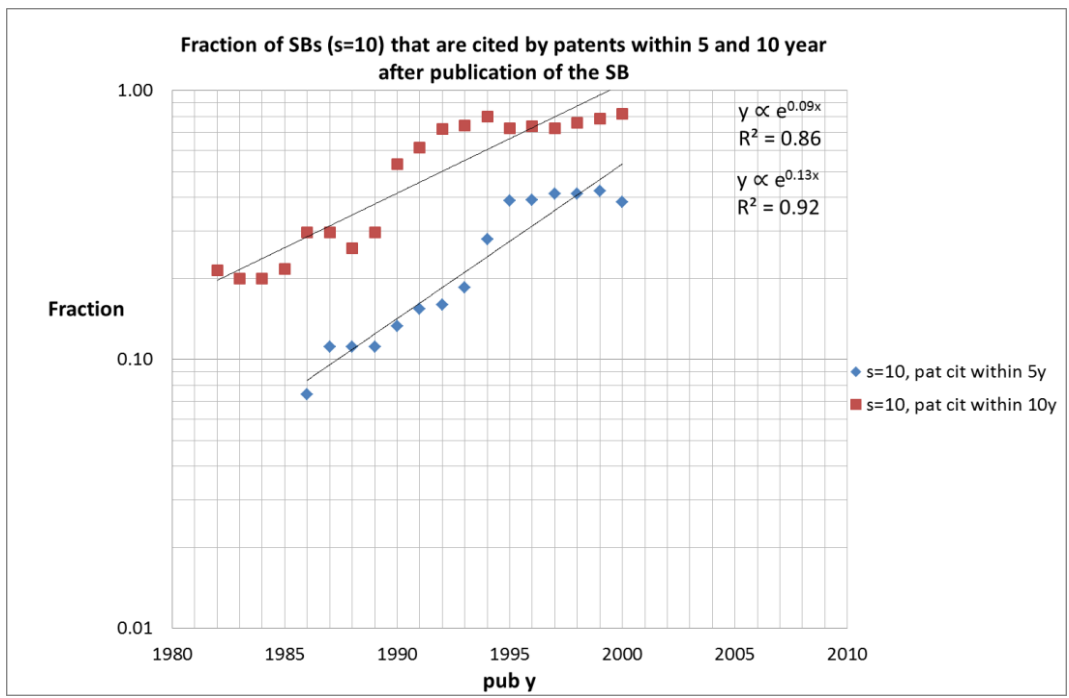

FIG. 15. Fraction of SBs with *s*=10 that are cited by patents within 5 and 10 year after publication of the SB.

To investigate the increasingly faster pace of technological impact in more detail, we analyzed the time lag between the filing year of the patent that is the *first* citer of the SB-SNPR and the publication year of the SB-SNPR (first patent citation year, **fpcy**). This *technological time lag* ranges for the SB-SNPRs from 1 to 30 years. We calculated averages of *fpcy* for successive, partly overlapping 5-year periods: 1980-1994, 1981-1985,.., 2004-2007. Fig.16 shows the results. We see that for SB-SNPRs with *s*=5 the technological time lag remains rather stable in the 1980's but after 1990 it becomes rapidly shorter (with about 7% per year) and around 2002 it becomes shorter than the sleeping time. We notice that this trend of the technological time lag does not differ that much for the SB-SNPRs with *s*=5 and *s*=10. The technological time lag becomes already in the early 1990's shorter than de sleeping time for the SB-SNPRs with *s*=10.

Just like our findings for the natural sciences and engineering these observations suggest that in the more recent years the probability that SBs are cited in a patent during sleeping time is increasing. This phenomenon is most related to two important developments: the increasing number of patents and the increasing number of patent citations to scientific publications. We intend to focus on these development in our follow-up research. In the next section we discuss our first findings.

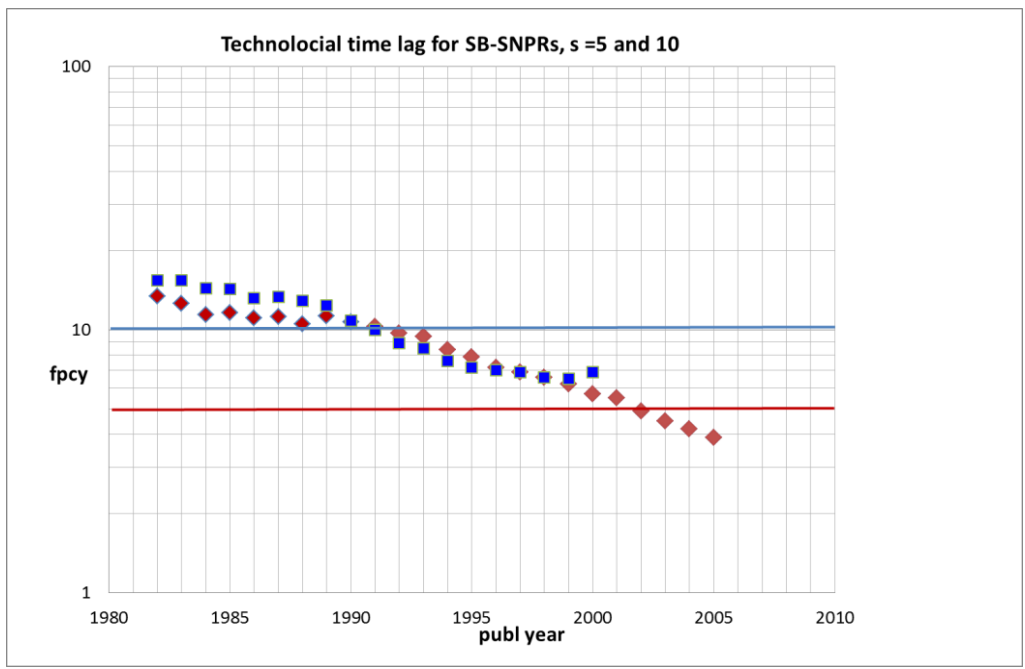

FIG. 16. Technological time lag for SB-SNPRs with *s*=5 (red diamonds) and *s*=10 (blue squares). In both cases *cs*(max)=1.0. The publication years on the abscissa are the middle years of the successive 5-years periods. The bold horizontal lines mark the lags (*fpcy*) of 5 (red line) and 10 years (blue line) which corresponds to the sleeping time for the SB-SNPRs with *s*=5 and *s*=10, respectively.

## Trends in patent scientific intensity

It hardly leads to doubt that the increase since 1980 of the number of SB-SNPRs that are cited by patents within 5 (for SBs with *s*=5) and 10 year (for SBs with *s*=10) after publication of the SB is influenced by the increasing number of patents as well as the increasing science intensity of patents, i.e., the share of patents citing scientific publications. For the sake of clarity we re-emphasize that these publications cited by patents are indicated by SNPRs, and if these publications are Sleeping Beauties we indicate them by SB-SNPRs.

To get more insight into the above discussed developments we first calculated several important general trends in patents. In Fig. 17 we show the trend of patents[3] as well as the trend of the number of patents that cite one or more scientific publications (i.e., patent with at least one SNPR). These trends are based on those patents that ultimately yield an EP or WO, but *also* a US patent publication. We think that these patents relate to inventions that are expected to be worth the most.

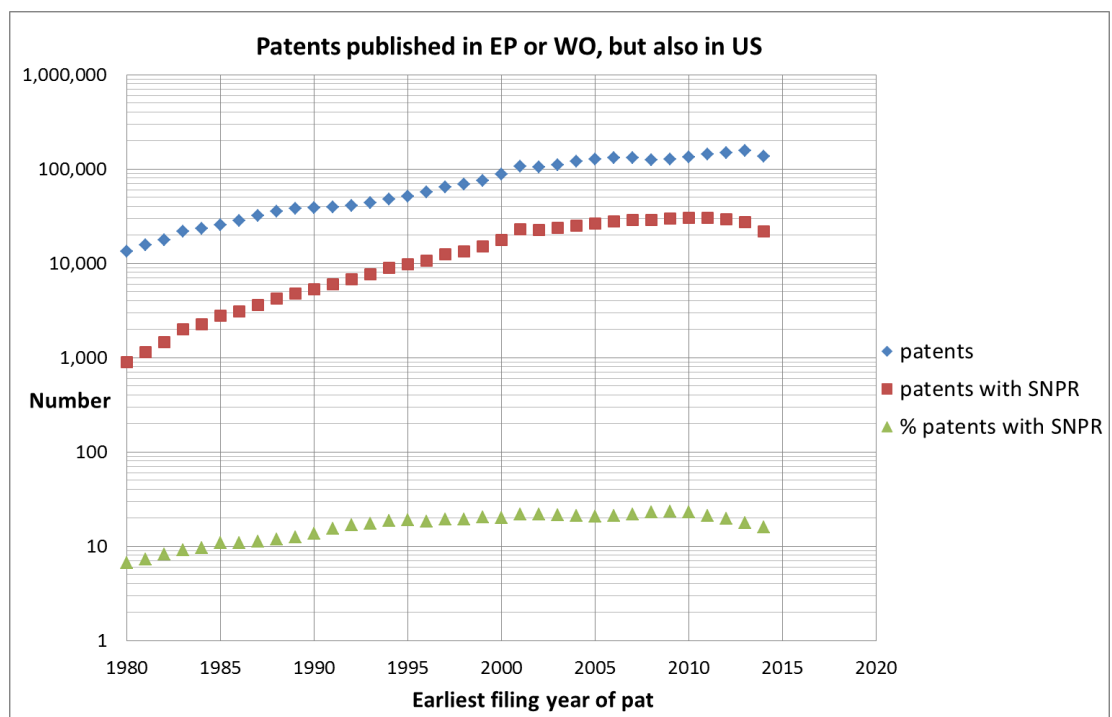

FIG. 17. Trend of the total number of patents with at least a patent publication in EP or WO but also in the US; the trend of patents with SNPRs, as well as the percentage of these patents.

As we notice in Fig. 17, data given more recently than 2015 are not shown. This is because for the recent years the patent database is not yet complete. The decrease from 2012 of the patents with SNPRs is probably caused by delays in citing SNPRs which may take several years. The percentage of patents with SNPRs, which we regard as discussed earlier as a simple indicator of patent science intensity, increased in the period 1980-1995 and becomes more or less stable since 1995 at around 20%. It is quite remarkable that the science intensity of patents remained stable in the last nearly 25 years. The decline in the most recent years is most probably again a database effect caused by the complexity of the patent procedures particularly the inclusion of references.

Next, we focus on the science intensity of technological fields that are relevant for our medical SB-SNPRs. In order to find these technological fields (defined by International Patent Classification (IPC) codes[4]), we determined the following variables: (a) the number of SB-SNPRs cited by patents in a specific technology field; (b) the share of technological fields in the total number of medical SB-SNPRs; (c) the share of a technology field in the total number of patents that have SNPRs (thus all SNPRs, not only the SB-SNPRs); and (d) the share of a technology field in all patents. We assume that variable (b) is the most interesting one for this study. We then calculated a composite indicator called 'science intensity index' by multiplying variable (b) (the share of technological fields in the total number of medical SB-SNPRs, we consider this as a weight of a technology field for the medical SB-SNPRs) with variable (c) (the share of a technology field in the total number of patents that have SNPR).

---

[3] As discussed earlier, with patents we mean patent families and particularly the member of a patent family with the earliest filing year. The patent families are defined in the DocDB database which is the internal patent database of the European Patent Database (EPO) that forms the basis for the PATSTAT database.
[4] Patents may have multiple IPC codes and thus they may belong to more than one technological field. The classification with 35 technology fields is explained by Schmoch (2008).

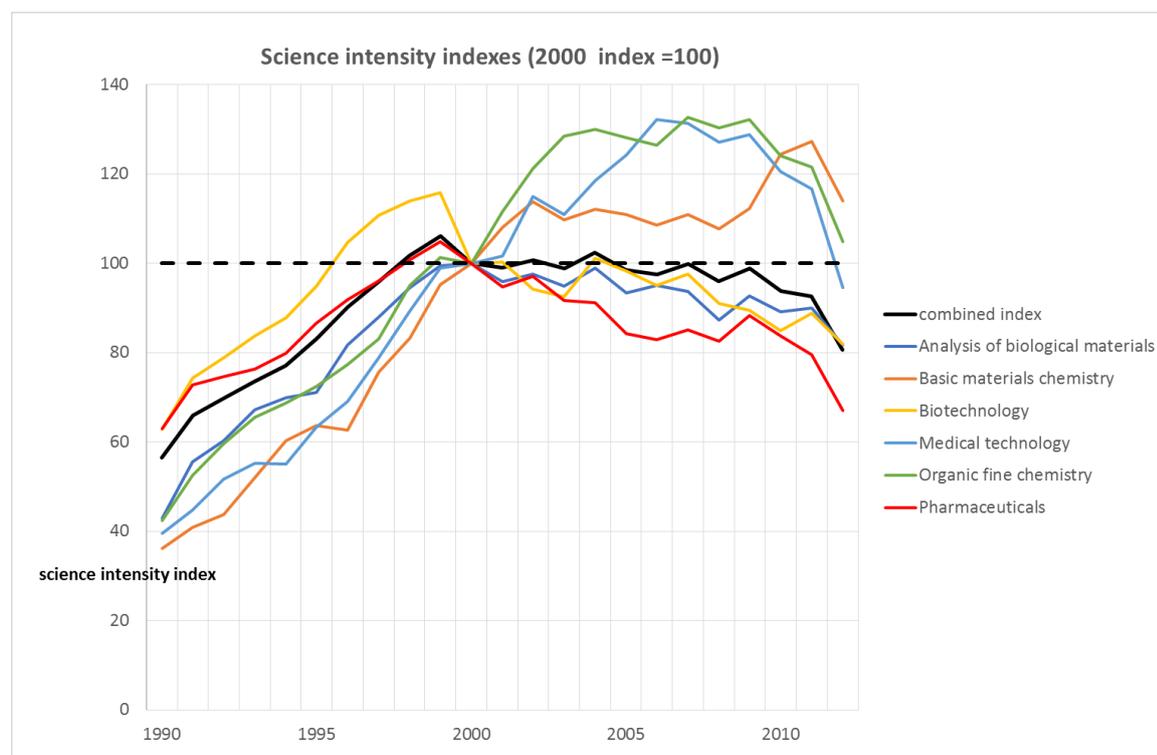

FIG. 18. Trend of the composite indicator for six technological fields that are the most important for medical SB-SNPRs. The indicator value in the year 2000 is taken as a reference value (index=100) for each field.

In Fig. 18 we show the trend of the science intensity index (1990-2012, years relate to the earliest filing year in a patent family[5]) for the six technological fields that are the most important for the medical SB-SNPRs: Pharmaceuticals, Medical Technology, Biotechnology, Organic Fine Chemistry, Analysis of Biological Materials, and Basic Materials Chemistry. We take 2000 as the index year (value =100). The science intensity index increases between 1990 and 2000, but from 2000 the six fields develop quite differently. However, these measurements must be improved by accurately taking into account the delay in patent citations to scientific publications. This will be the subject of a follow-up study and therefore proper normalization of the number of SB-SNPRs on the basis of the science intensity is in the context of this study not possible.

---

[5] The earliest filing date is the well-documented date closest to the time of invention and can therefore be used as a proxy for that moment of time.

## Influence on the technological time lag by inventor-author self-citation

In foregoing studies (van Raan 2017a; van Raan & Winnink 2018) focusing on physics, chemistry, engineering and computer science, we investigated the extent to which at least one of the *inventors of patents citing* SB-SNPRs is also an *author of the cited SB-SNPR*. We call this inventor-author self-citation (Noyons et al, 1994). We concluded that only for a small minority (5%) of the SB-SNPRs the authors are also inventors of the technology described in the citing patent. In this study we find that for the 663 SB-SNPRs with *s*=5 inventor-author self-citation occurs in the case of 59 SB-SNPRs, i.e., 9%. These 59 SB-SNPRs are cited by 206 patents, of which nearly half, 95, deal with inventor-author self-citation. This occurrence of inventor-author self-citation is comparable though somewhat higher as compared to our findings for physics, chemistry, engineering and computer science.

A new element in this study is the time trend of inventor-author self-citations. In Fig.19 we show for successive five-year time blocks the number of SB-SNPRs with *s*=5 of which at least one of the authors is also one of the inventors of the citing patent. This number increases with about 11% per year: in recent years the occurrence of inventor-author self-citations is about four to five times higher than in the 1980's.

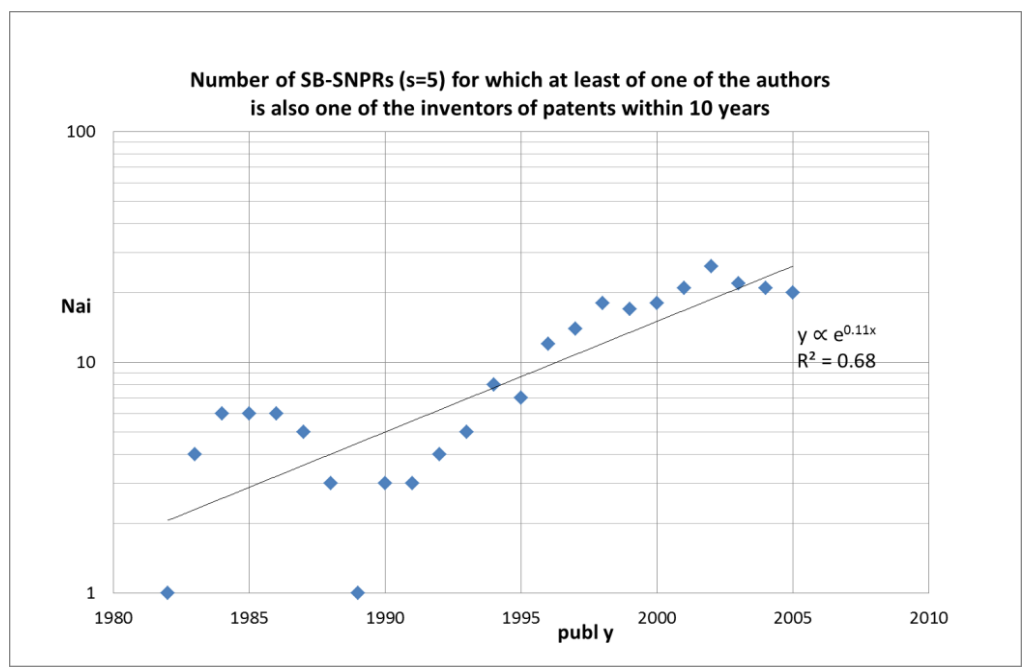

FIG. 19. Number of SB-SNPRs (*s*=5) of which at least one of the authors is also one of the inventors of the citing patent (**Nai**) as a function of the publication year of the SB-SNPRs.

Are inventor-author self-citations responsible for earlier first patent citations of the Sleeping Beauties and thus shorter technological time lags? We used the analysis of the technological time lag discussed in the forgoing section to investigate this. For SB-SNPRs with *s*=5 we removed in our analysis from 1990 the patents that are the first citers in the case of inventor-author self-citation (inventor-author correction). The results are shown in Fig.20. The data indicated with the blue triangles are the corrected data. Without corrections for inventor-author self-citation the results are the same as in Fig.14 (from 1990). We see that inventor-author self-citations indeed result in shorter technological time lags, but this effect is small. This finding confirm the results of our earlier studies that inventor-author self-citation is quite rare. In these earlier studies we also found on the basis of a number of individual cases earlier *scientific* awakening triggered by inventor-author self-citation is small. In this study we find that earlier *technological* awakening triggered by inventor-author self-citation is also small.

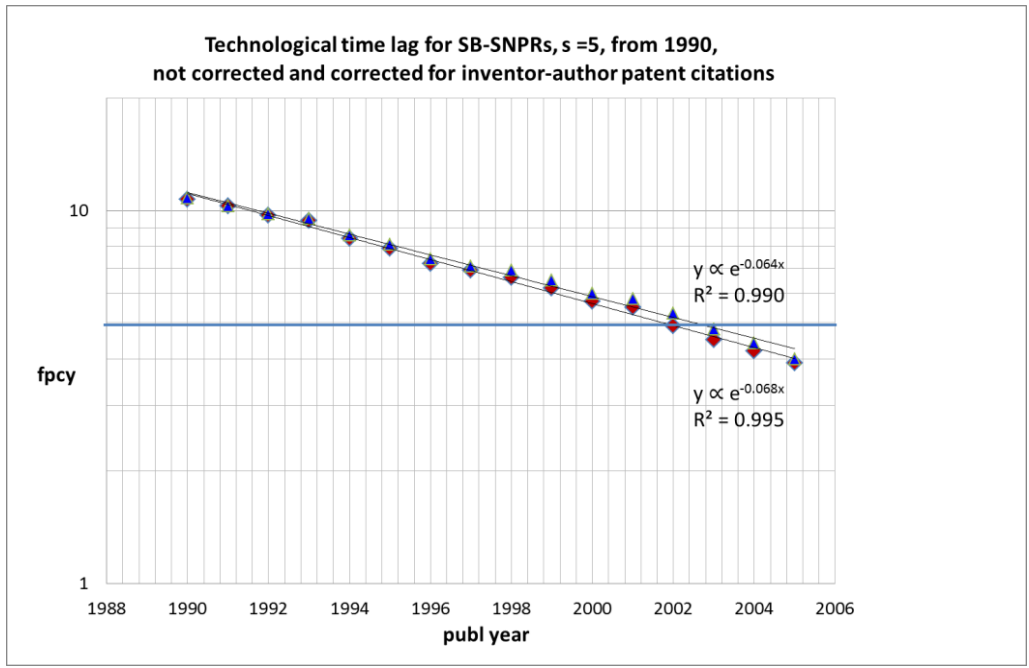

FIG. 20. Technological time lag for SB-SNPRs with $s$=5: not-corrected (red diamonds) and corrected (blue triangles) for inventor-author patent citations. The years indicated on the abscissa are the middle years of the successive 5-years periods. The bold horizontal line indicates a time lag of 5 years which is the sleeping time for the SB-SNPRs.

## Scientific and technological impact

Just as in the case of publications, also patents show a wide variety of impact. Only a relatively small amount of patents represents important technological breakthroughs (Albert et al. 1991). Patent-to-patent citations provide a first indication of the importance of the cited patents (Trajtenberg 1990), particularly if they are highly cited. Harhoff et al. (1999) found that patents renewed to full-term (which is the maximum duration of the patent protection, mostly 20 years) were significantly more highly cited than patents allowed to expire before their full term. The higher an invention's economic value estimate was, the more the patent was subsequently cited. For an overview of patent citation analysis studies we refer to van Raan (2017b).

We first analyzed for all patents citing the SB-SNPRs the number of times they themselves are cited by other patents. In order to analyze a time trend and to compare the successive years within our measuring period 1980-2007, we have to work with fixed citation windows. First, we select all patents that cite SB-SNPRs ($s$=5) *within 10 years after the publication of the SB*. Second, for these patents we counted the citations they receive from other patents *within 5 year after of the filing date of the cited patent*. By ranking these patents by the number of their patent-citations we were able to identify the top-20% cited patents. Finally we determine the number of SB-SNPRs that are cited by one or more of these top-20% patents. We find that of the 663 SB-SNPRs with $s$=5, 107 (16%) are cited by one of more of the 381 top-20% highly cited patents. The last SB publication year we can take into account is 2002: we need a 10-year period for the SBs to be cited by a patent after their publication, and subsequently 5 years for these patent to be cited by other patents, so in total 15 years.

The last complete year of our patent database is 2015 and thus the last SB publication year must be 2002. Fig.21 shows that, remarkably, the fraction of these SB-SNPRs with citing patents belonging to the top-20% remains more or less constant around 0.16.

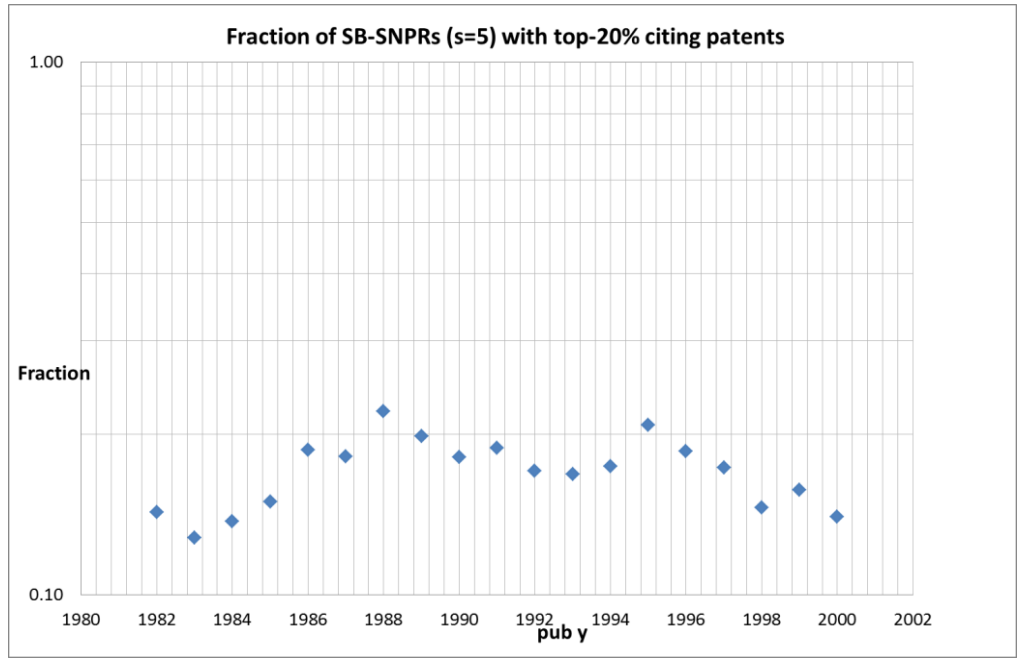

FIG. 21. Fraction of SB-SNPRs (s=5) that are cited by top-20% patents

For both the SB-SNPRs as well as the SB-nonSNPRs with $s$=5 we found no significant correlation between the number of citations by other publications during the sleeping period (measured by $cs$) and the number of citations by other publications during the awake period (measured by $ca$). In other words, similar to our earlier findings for the natural sciences and engineering, the depth of the sleep is no predictor for the awake intensity. Also we did not find a significant correlation between the awake intensity and the impact of patents that cite the SB-SNPR, i.e., the number of time these patents are cited themselves by other patents within ten tear after the filing date of the patent. This means that the scientific impact of Sleeping Beauties is generally not related to the technological importance of the SBs, as far as measured with number and impact of the citing patents. Again this is similar to our earlier findings for the natural sciences and engineering. Remarkably, however, we do find that the average number of citations (by other publications) during the awakening period ($ca$) tends to be higher for the SB-SNPRs ($s$=5), but this difference disappears gradually, see Fig.22.

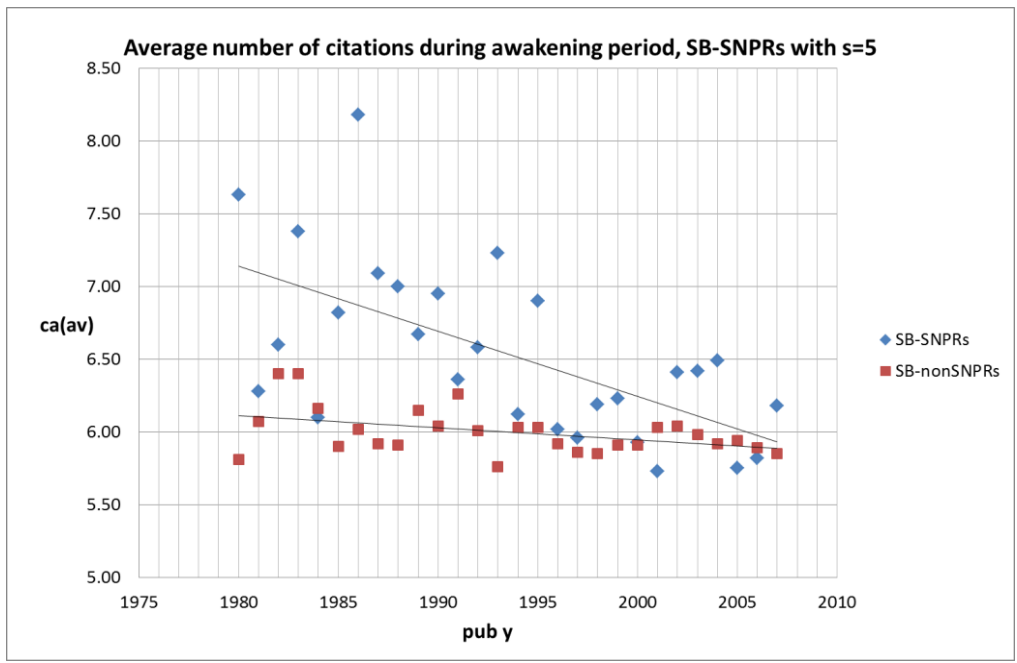

FIG. 22. Awake citation-intensity for SN=SNPRs and SB-nonSNPRs with $s$=5.

In our set of 5,247 SBs with $s$=5 one SB, also an SB-SNPR, immediately strikes the eye because of an enormous increase of citations: Matthews et al (1985, University of Oxford), see Figs. 23 and 24 which show the sleeping period followed by an exponential increase of citations. Fig. 24 also compares the citation trend of the Matthews SB-SNPR with an average SB-SNPR published in 1985. The citation rate of this SB-SNPR during sleeping time was $cs$=1.0 ('deep sleep') and the citation rate during the awake period was $ca$=7.0 which is just slightly above the average of all 5,261 SBs with $s$=5. But at the end of the awake period an exponential increase of citations starts. From 2008 until now the paper has more than 1,000 citations per year, with a total of 18,141 (WoS Core Collection, August 14, 2018) which makes this SB-SNPR currently the third highest cited paper of all WoS-covered papers published in 1985 and places this SB-SNPR within the top-50 highest cited papers ever.

This SB-SNPR of Matthews and co-authors published in the journal Diabetologia deals with a mathematical, computerized model to determine plasma glucose and insulin concentrations by their interaction in a feedback loop. This model enables to accurately predict the homeostatic concentrations which arise from varying degrees of cell function deficiency and insulin resistance. The five main research fields of the citing papers are endocrinology & metabolism (38%), nutrition dietetics (13%), general internal medicine (6%), cardiac & cardiovascular systems (6%), peripheral vascular diseases (6%). The journal Diabetologia ranks 16 by journal impact in the field Endocrinology & Metabolism. So we see that a very highly cited paper is not necessarily a paper in a top-journal in terms of, say, the top-5 in journal impact.

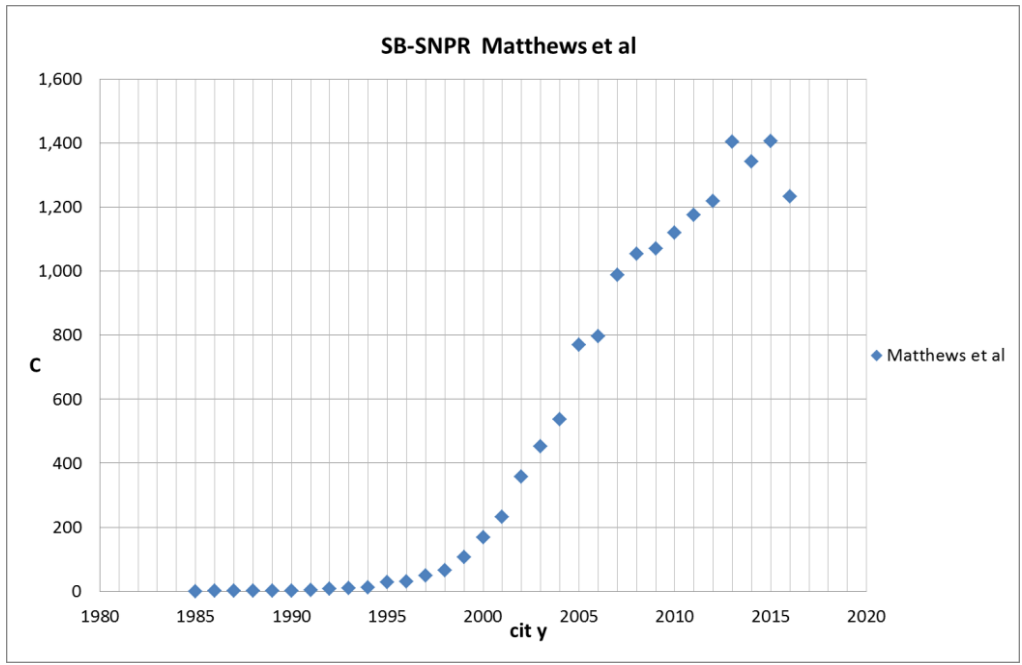

FIG. 23. Trend in the number of citations received by the Matthews et al, an SB-SNPR published in 1985. On the abscissa the years of citation are given, on the ordinate the number of citations.

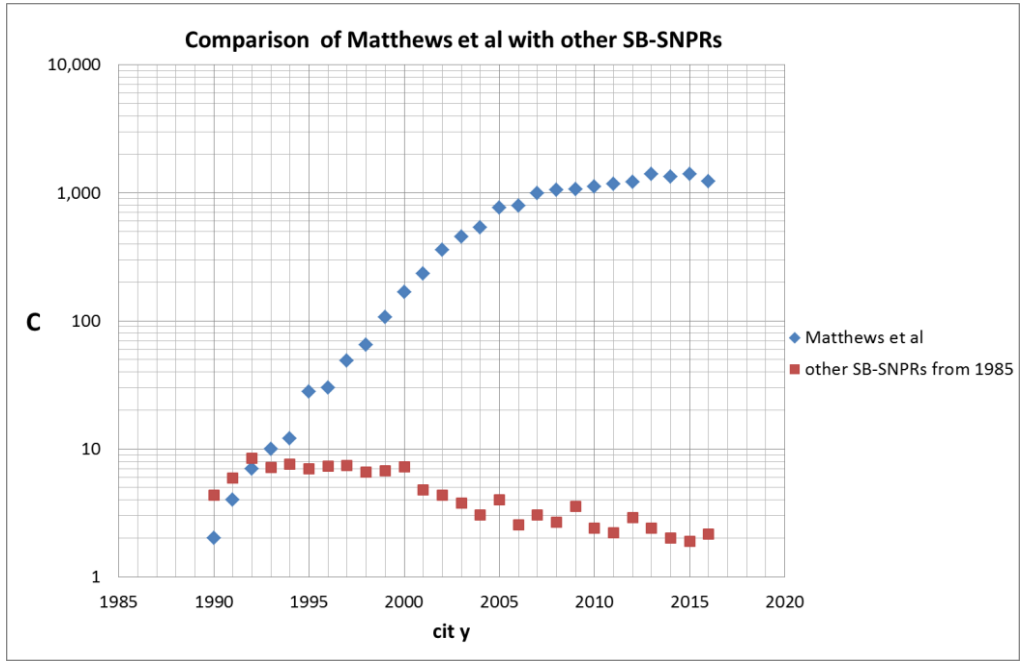

FIG. 24. Trend in the number of citations received by the Matthews et al compared to the average of the other SB-SNPRs published in 1985. On the abscissa the years of citation are given, on the ordinate the number of citations.

From the 18,141 papers that cite the Matthews SB-SNPR we selected the 500 most cited of which we made a co-citation analysis and mapping with the CWTS VoSviewer (van Eck and Waltman 2010). For a detailed discussion of the use of the mapping tool VoSviewer in our Sleeping Beauties research we refer to our earlier paper (van Raan 2015).

The results are shown in Fig.25. The co-citation clusters of the references of the top-500 citing papers are from a variety of medical fields which is nicely reflected by the clusters in the butterfly-like structures with different colors. The green cluster represents research on the metabolic syndrome and cardiovascular diseases; the purple cluster relates to insulin resistance, obesity and the role of hormones; the blue cluster also focuses on insulin resistance but particularly liver diseases; the red cluster focuses on predicting and assessing insulin resistance; and the yellow cluster relates to genetic studies of diabetes. Undoubtedly the Matthews SB-SNPR is the most central paper as it is, by definition, cited by all citing papers.

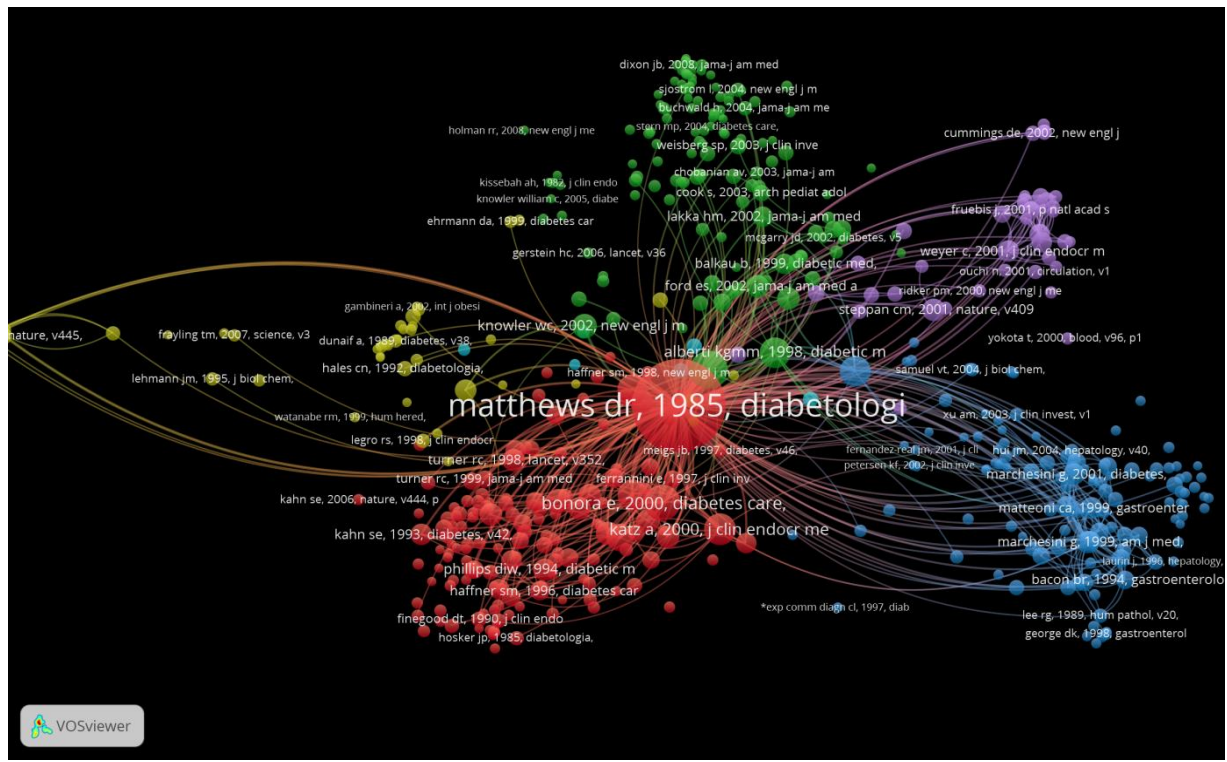

FIG. 25. Co-citation map of the 500 most highly cited papers that cite the Matthews et al SB-SNPR (co-citation threshold=3).

The VoSviewer also enables us to measure all co-occurrences of any possible pair of concepts in a set of papers. In this way we created a map in which the conceptual structure of the research represented by the set of the top-500 papers citing the Matthew SB-SNPR. The results are shown in Fig. 26. We clearly see concept clusters representing different themes. Major clusters are the blue one around glucose, insulin resistance, and obesity; the green cluster around diabetes, beta-cell function and homeostasis; the red cluster around cardiovascular diseases. Clearly insulin resistance plays a central role.

FIG. 26. Concept co-occurrence map of the 500 most highly cited papers that cite the Matthews et al SB-SNPR (co-occurrence threshold=5).

Even within 10 years after the publication of the Matthews SB-SNPR there are no citing patents. The first patent citation is in 2001, 16 years after the publication of this SB-SNPR. As we have seen in Fig. 23, the number of (scientific) citations has already started to rise. It takes until 2010 before the number of citing patents increases rapidly, in the PatStat database Spring version 2016 39 citing patents are registered, several of these patents are highly cited by other patents. The patents are from the following technology fields: pharmaceuticals, organic fine chemistry, basic materials chemistry, biotechnology, analysis of biological materials, medical technology. Notice the difference between the citing scientific and the citing technological fields. Perhaps surprisingly, the patents fields most clearly show where the Matthews SB-SNPR is about, particularly analysis of biological materials; the scientific fields show where the work of Matthews et al is applied.

## Conclusions

In this paper we investigate in the medical research fields recent Sleeping Beauties with a special focus on those SBs that are cited in patents (SB-SNPRs). We find that the increasing trend of the relative number of SBs comes to an end around 1998. It confirms our earlier observations in the natural science and engineering fields and supports our conjecture that the expanding worldwide facilities to access scientific publications seems to have stopped increasing trends in the occurrence of SBs. Apparently, however, this does not prevent that a more or less constant fraction of publications still becomes an SB.

We measured for the entire period 1980-2007 the scaling of the number of SBs with sleeping period length, with during-sleep citation-intensity and with the awake citation-intensity. Particularly the scaling with sleeping period length shows a remarkable time-dependent change: during our measuring period the scaling exponent doubled. Our explanation of this change is that in the early 1980's the probability to have SBs with very long sleeping times was considerably higher than it is nowadays. On the basis of these scaling measurements we determined the *Grand Sleeping Beauty Equation*. This equation describes important quantitative characteristics of SBs. If a publication is *twice as longer in deep sleep*, the probability that it awakes with, on average, at least five citation per year during five years, is about an *order of magnitude less*. In other words, the longer the sleeping period, the less probable it is that a publication will awake. Indeed a finding that we can intuitively recognize. Next, the probability to find a publication that has twice as much citations in deep sleep (e.g., 4 versus 2 in five years sleep), is about factor 2.5 higher. In other words, the less deep the sleep, the more SBs will be found, because we are moving toward 'normal' publications. The probability for a higher awake citation-intensity decreases very rapidly. For instance, the probability that a SB will have during the awake period a citation intensity twice as large as another SB (e.g., 10 versus 5 citations per year in the five years awake period), is about a factor 50 lower.

Similar to citations given by publications, also the number of citations given to SBs by patents is characterized by a skewed distribution. Our study demonstrates that for SB-SNPRs with *s*=5 the fraction for at least one patent citation within 5 years after publication, and thus within their sleeping period, is in the 1980's about 25%. But about 20 years later, in the first decade of this century, the situation has changed drastically: 70-80% of SB-SNPRs are cited within their sleeping period. For SB-SNPRs with the longer sleep period *s*=10 the fraction for at least one patent citation within 5 years after publication is in the 1980's about 10% and in the first decade of this century it is 40%, again a substantial change. It turns out that the fraction of SB-SNPRs that are cited by patents *within their sleeping period* is exponentially increasing and after the year 2000 practically all SB-SNPRS are cited by patents within their sleeping period.

To investigate the increasingly faster pace of technological impact in more detail, we analyzed the time lag between the filing year of the patent that is the *first* citer of the SB-SNPR and the publication year of the SB-SNPR. This *technological time lag* ranges for the SB-SNPRs from 1 to 30 years. We find that for SB-SNPRs with *s*=5 the technological time lag remains rather stable in the 1980's but after 1990 it becomes rapidly shorter (with about 7% per year) and around 2002 it becomes shorter than the sleeping time. The technological time lag becomes already in the early 1990's shorter than de sleeping time for the SB-SNPRs with *s*=10. This can be expected given the similarity of the trend of the technological time lag for both the SB-SNPRs with *s*=5 as well as with *s*=10. Like our earlier findings for the natural sciences and engineering these observations again suggest that, on average, in the more recent years the majority of SB-SNPRs are cited by one or more patents before the 'scientific wakening'.

A new element in this study is the time trend of inventor-author self-citations. With inventor-author self-citations we mean that at least one of the *inventors of patents citing* SB-SNPRs is also an *author of the cited SB-SNPR*. This number increases (SB-SNPRs with *s*=5) with about 11% per year: in recent years the occurrence of inventor-author self-citations is about four to five times higher than in the 1980's. We find that inventor-author self-citations may result in shorter technological time lags, but this effect is small. The findings in this study on inventor-author self-citations in medical SB-SNPRs confirm the results of our earlier studies in the natural sciences and engineering that inventor-author self-citation is quite rare. In these earlier studies we also found on the basis of a number of individual cases earlier *scientific* awakening triggered by inventor-author self-citation is small. In this study we find that earlier *technological* awakening triggered by inventor-author self-citation is also small.

We also studied the technological impact of the patents citing SB-SNPRs by analyzing the extent to which these patents themselves are cited later on by other patents within ten years after the filing date of the patent. This enabled us to identify the top-20% highly cited patents. Remarkably, the fraction of SB-SNPRs with citing patents belonging to the top-20% remains more or less constant over the entire measuring period.

For both the SB-SNPRs as well as the SB-nonSNPRs with $s$=5 we found no significant correlation between the number of citations by other publications during the sleeping period and the number of citations by other publications during the awake period. In other words, the depth of the sleep is no predictor for the awake intensity. Remarkably, however, we do find that the average number of citations (by other publications) during the awakening period tends to be higher for the SB-SNPRs ($s$=5), but this difference disappears gradually. We also did not find a significant correlation between the awake intensity and the technological impact of patents that cite the SB-SNPR. This means that the scientific impact of Sleeping Beauties is generally not related to the technological importance of the SBs, as far as measured with number and impact of the citing patents. Again this is similar to our earlier findings for the natural sciences and engineering.

Within our set of medical Sleeping Beauties one SB stands out, the Matthews SB-SNPR, published in 1985, on a mathematical, computerized model to determine plasma glucose and insulin concentrations, a method which turned out to be of great significance for diabetes patients. Even within 10 years after the publication of this SB-SNPR there are no citing patents. The first patent citation is in 2001, 16 years after the publication of this SB-SNPR. It takes until 2010 before the number of citing patents and with that the *technological impact* of this SB-SNPR increases rapidly, in the PatStat database Spring version 2017 39 citing patents are registered, several of these patents are highly cited by other patents. But most striking is the delayed but enormously increasing *scientific* impact of this SB-SNPR: it took more than five years before the Matthews SB-SNPR started to become reasonably cited (more than 5 citations per year), then the number of citations increased exponentially and from 2008 until now the paper has more than 1,000 citations per year, with a total of nearly 20,000 which makes this SB-SNPR currently the third highest cited paper of all WoS-covered papers published in 1985 and places this SB-SNPR within the top-50 highest cited papers ever.

## Keywords

Sleeping Beauties, medical research, patent citations, technological time lag, inventor-author relations, technological impact, technological awakening, scientific awakening.

## References

Albert, M.B., Avery, D., Narin, F. & McAllister, P. (1991). Direct Validation of Citation Counts as Indicators of Industrially Important Patents. *Research Policy* 20(3):251-259.

Callaert, J., Pellens, M. & van Looy, B. (2014). Sources of Inspiration? Making Sense of Scientific References in Patents. Scientometrics 98(3): 1617-1629.

Garfield, E. (1970). Would Mendel's work have been ignored if the Science Citation Index was available 100 years ago? *Essays of an Information Scientist* 1, 69–70; also in *Current Contents* 2, 69-70.

Garfield, E. (1980). Premature discovery or delayed recognition –why? *Essays of an Information Scientist* 4, 488–493; also in *Current Contents* 21, 5-10, 1980.

Garfield, E. (1989). Delayed recognition in scientific discovery: Citation frequency analysis aids the search for case histories. *Current Contents* 23, 3-9.

Garfield, E. (1990) More delayed recognition. Part 2. From Inhibin to Scanning electron microscopy. *Essays of an Information Scientist* 13, 68–74; also in Current Contents 9, 3-9, 1990.

Harhoff, D., Narin, F., Scherer, M. & Vopel, K. (1999). Citation Frequency and the Value of Patented Inventions. *Review of Economics and Statistics* 81(3):511-515.

Matthews, D.R., Hosker, J.P., Rudenski A.S., Naylor, B.A., Treacher, D.F. & Turner, R.C. (1985). Homeostasis Model Assessment - Insulin Resistance and Beta-Cell Function from Fasting Plasma-Glucose and Insulin Concentrations in Man. Diabetologia 28(7): 412-419.

Narin, F., Hamilton, K.S. & Olivastro, D. (1997). The increasing linkage between US technology and public science. *Research Policy* 26, 317-330.

Noyons, E.C.M., van Raan, A.F.J., Grupp, H. & Schmoch, U. (1994). Exploring the science and technology interface: inventor-author relations in laser medicine research. *Research Policy* 23(4):443-457.

Stent, G.S. (1972). Prematurity and uniqueness in scientific discovery. *Scientific American* 227(6) 84–93.

Trajtenberg, M. (1990). A Penny for Your Quotes: Patent Citations and the Value of Innovations. *RAND Journal of Economics* 21(1):172-187.

van Eck N.J. & Waltman, L. (2010). Software survey: VOSviewer, a computer program for bibliometric mapping. *Scientometrics* 84(2), 523–538. For more information see http://www.vosviewer.com/Home.

Schmoch U. (2008). Concept of a Technology Classification for Country Comparisons, see http://www.wipo.int/edocs/mdocs/classifications/en/ipc_ce_41/ipc_ce_41_5-annex1.pdf .

van Raan, A.F.J. (2004). Sleeping Beauties in Science. *Scientometrics* 59 (3), 461-466.

van Raan, A.F.J. (2015). Dormitory of Physical and Engineering Sciences: Sleeping Beauties May Be Sleeping Innovations. *PLoS ONE* 10(10): e0139786.

van Raan, A.F.J. (2017a). Sleeping Beauties Cited in Patents: Is there also a Dormitory of Inventions? *Scientometrics* 110(3) 1123-1156.

van Raan, A.F.J. (2017b). Patent Citations Analysis and Its Value in Research Evaluation: A Review and a New Approach to Map Technology-relevant Research. *Journal of Data and Information Science* 2(1) 13–50.

van Raan, A.F.J. & Winnink, J.J. (2018). Do younger Sleeping Beauties prefer a technological prince? Scientometrics 114: 701-717.

Winnink, J.J. & Tijssen, R.J.W. (2014). R&D dynamics and scientific breakthroughs in HIV/AIDS drugs development: the case of integrase inhibitors. *Scientometrics* 101(1): 1–16.

# Supplementary Material

TABLE S1. WoS Fields codes and names of medical research fields

| Clinical Medicine | |
|---|---|
| **WoS field code and name** | |
| 9 | ALLERGY |
| 10 | ANATOMY & MORPHOLOGY |
| 11 | ANDROLOGY |
| 12 | ANESTHESIOLOGY |
| 32 | ONCOLOGY |
| 33 | CARDIAC & CARDIOVASCULAR SYSTEMS |
| 56 | EMERGENCY MEDICINE |
| 61 | DENTISTRY/ORAL SURGERY & MEDICINE |
| 62 | DERMATOLOGY |
| 74 | ENDOCRINOLOGY & METABOLISM |
| 99 | GASTROENTEROLOGY & HEPATOLOGY |
| 100 | GENETICS & HEREDITY |
| 105 | GERIATRICS & GERONTOLOGY |
| 106 | HEALTH POLICY & SERVICES |
| 107 | HEMATOLOGY |
| 113 | PUBLIC, ENVIRONMENTAL & OCCUPATIONAL HEALTH |
| 114 | IMMUNOLOGY |
| 116 | INFECTIOUS DISEASES |
| 122 | MEDICINE, LEGAL |
| 143 | MEDICAL INFORMATICS |
| 146 | MEDICINE, GENERAL & INTERNAL |
| 147 | METALLURGY & METALLURGICAL ENGINEERING |
| 148 | MEDICINE, RESEARCH & EXPERIMENTAL |
| 166 | CLINICAL NEUROLOGY |
| 167 | NEUROSCIENCES |
| 169 | NURSING |
| 170 | NUTRITION & DIETETICS |
| 171 | OBSTETRICS & GYNECOLOGY |
| 174 | OPHTHALMOLOGY |
| 177 | ORTHOPEDICS |
| 178 | OTORHINOLARYNGOLOGY |
| 180 | PARASITOLOGY |
| 181 | PATHOLOGY |
| 182 | PEDIATRICS |
| 183 | PHARMACOLOGY & PHARMACY |
| 199 | PSYCHIATRY |
| 206 | RADIOLOGY, NUCLEAR MEDICINE & MEDICAL IMAGING |
| 207 | REHABILITATION |
| 208 | RESPIRATORY SYSTEM |
| 209 | REPRODUCTIVE BIOLOGY |
| 210 | RHEUMATOLOGY |

TABLE S2. Numbers of the identified SBs

| s=5 | cs(max) | | | | |
|---|---|---|---|---|---|
| | 0.2 | 0.4 | 0.6 | 0.8 | 1.0 |
| 1980 | 3 | 6 | 15 | 28 | 52 |
| 1981 | 5 | 11 | 18 | 32 | 55 |
| 1982 | 1 | 6 | 14 | 33 | 56 |
| 1983 | 4 | 10 | 19 | 35 | 66 |
| 1984 | 3 | 7 | 22 | 35 | 62 |
| 1985 | 2 | 9 | 27 | 37 | 73 |
| 1986 | 8 | 18 | 31 | 50 | 87 |
| 1987 | 4 | 13 | 41 | 76 | 122 |
| 1988 | 5 | 20 | 44 | 78 | 133 |
| 1989 | 10 | 28 | 55 | 107 | 176 |
| 1990 | 14 | 33 | 66 | 127 | 213 |
| 1991 | 21 | 49 | 95 | 160 | 252 |
| 1992 | 15 | 30 | 62 | 109 | 196 |
| 1993 | 8 | 22 | 47 | 91 | 152 |
| 1994 | 6 | 16 | 48 | 75 | 123 |
| 1995 | 3 | 13 | 42 | 84 | 140 |
| 1996 | 14 | 27 | 52 | 93 | 168 |
| 1997 | 10 | 28 | 57 | 108 | 211 |
| 1998 | 14 | 37 | 76 | 127 | 224 |
| 1999 | 15 | 47 | 98 | 185 | 300 |
| 2000 | 14 | 38 | 95 | 208 | 343 |
| 2001 | 13 | 36 | 91 | 194 | 350 |
| 2002 | 11 | 44 | 100 | 190 | 351 |
| 2003 | 6 | 27 | 78 | 169 | 316 |
| 2004 | 10 | 32 | 78 | 169 | 302 |
| 2005 | 12 | 27 | 52 | 113 | 249 |
| 2006 | 7 | 24 | 62 | 124 | 231 |
| 2007 | 11 | 25 | 64 | 125 | 244 |

| s=10 | cs(max) | | | | |
|---|---|---|---|---|---|
| | 0.2 | 0.4 | 0.6 | 0.8 | 1.0 |
| 1980 | 0 | 0 | 5 | 8 | 13 |
| 1981 | 0 | 0 | 1 | 2 | 9 |
| 1982 | 0 | 0 | 3 | 4 | 10 |
| 1983 | 1 | 5 | 6 | 8 | 14 |
| 1984 | 0 | 3 | 4 | 13 | 29 |
| 1985 | 3 | 7 | 15 | 23 | 38 |
| 1986 | 0 | 6 | 14 | 25 | 38 |
| 1987 | 0 | 2 | 5 | 11 | 25 |
| 1988 | 0 | 0 | 4 | 8 | 25 |
| 1989 | 1 | 1 | 1 | 5 | 17 |
| 1990 | 1 | 2 | 9 | 13 | 18 |
| 1991 | 0 | 3 | 7 | 11 | 17 |
| 1992 | 0 | 3 | 5 | 12 | 23 |
| 1993 | 1 | 2 | 2 | 4 | 12 |
| 1994 | 0 | 2 | 3 | 9 | 24 |
| 1995 | 0 | 3 | 7 | 17 | 31 |
| 1996 | 0 | 1 | 3 | 13 | 34 |
| 1997 | 1 | 3 | 11 | 24 | 43 |
| 1998 | 2 | 3 | 5 | 21 | 46 |
| 1999 | 0 | 5 | 10 | 23 | 48 |
| 2000 | 1 | 4 | 8 | 21 | 43 |
| 2001 | 1 | 2 | 5 | 15 | 32 |
| 2002 | 3 | 4 | 7 | 15 | 25 |

| | cs(max) | | | | |
|---|---|---|---|---|---|
| s=15 | 0.2 | 0.4 | 0.6 | 0.8 | 1.0 |
| 1980 | 1 | 1 | 1 | 1 | 7 |
| 1981 | 0 | 1 | 1 | 5 | 10 |
| 1982 | 0 | 0 | 2 | 11 | 15 |
| 1983 | 2 | 2 | 2 | 5 | 12 |
| 1984 | 0 | 0 | 1 | 2 | 5 |
| 1985 | 2 | 2 | 3 | 4 | 7 |
| 1986 | 0 | 0 | 2 | 6 | 8 |
| 1987 | 1 | 2 | 2 | 4 | 7 |
| 1988 | 0 | 0 | 1 | 4 | 8 |
| 1989 | 0 | 1 | 2 | 8 | 11 |
| 1990 | 0 | 0 | 2 | 6 | 14 |
| 1991 | 0 | 0 | 5 | 7 | 13 |
| 1992 | 0 | 0 | 1 | 5 | 11 |
| 1993 | 0 | 0 | 2 | 4 | 10 |
| 1994 | 0 | 2 | 4 | 7 | 11 |
| 1995 | 1 | 3 | 6 | 10 | 17 |
| 1996 | 0 | 2 | 5 | 7 | 17 |
| 1997 | 1 | 1 | 4 | 10 | 16 |

| | cs(max) | | | | |
|---|---|---|---|---|---|
| s=20 | 0.2 | 0.4 | 0.6 | 0.8 | 1.0 |
| 1980 | 0 | 0 | 1 | 2 | 3 |
| 1981 | 0 | 0 | 1 | 3 | 4 |
| 1982 | 0 | 0 | 1 | 2 | 4 |
| 1983 | 1 | 1 | 2 | 3 | 5 |
| 1984 | 0 | 1 | 1 | 4 | 10 |
| 1985 | 0 | 1 | 4 | 6 | 11 |
| 1986 | 0 | 0 | 0 | 3 | 6 |
| 1987 | 0 | 1 | 1 | 5 | 11 |
| 1988 | 0 | 0 | 1 | 5 | 10 |
| 1989 | 0 | 2 | 3 | 4 | 14 |
| 1990 | 2 | 3 | 3 | 5 | 9 |
| 1991 | 1 | 1 | 3 | 7 | 12 |
| 1992 | 0 | 1 | 2 | 7 | 11 |

| | cs(max) | | | | |
|---|---|---|---|---|---|
| s=25 | 0.2 | 0.4 | 0.6 | 0.8 | 1.0 |
| 1980 | 0 | 3 | 3 | 6 | 11 |
| 1981 | 0 | 0 | 0 | 1 | 2 |
| 1982 | 0 | 2 | 4 | 5 | 8 |
| 1983 | 0 | 1 | 1 | 4 | 6 |
| 1984 | 1 | 2 | 3 | 5 | 8 |
| 1985 | 0 | 1 | 4 | 7 | 10 |
| 1986 | 0 | 1 | 3 | 5 | 11 |
| 1987 | 0 | 0 | 1 | 3 | 6 |

| | cs(max) | | | | |
|---|---|---|---|---|---|
| s=30 | 0.2 | 0.4 | 0.6 | 0.8 | 1.0 |
| 1980 | 0 | 0 | 0 | 2 | 6 |
| 1981 | 0 | 0 | 2 | 3 | 9 |
| 1982 | 0 | 1 | 2 | 4 | 4 |

TABLE S3. Trend of real and normalized number of SBs with $s$=5 ($cs$(max)=1) in the medical research fields, numbers are based on successive five-years blocks.

| $s$=5 | not-normalized | normalized |
|---|---|---|
| 1980-84 | 291 | 520 |
| 1981-85 | 312 | 529 |
| 1982-86 | 344 | 564 |
| 1983-87 | 410 | 651 |
| 1984-88 | 477 | 734 |
| 1985-89 | 591 | 869 |
| 1986-90 | 731 | 1,044 |
| 1987-91 | 896 | 1,244 |
| 1988-92 | 970 | 1,311 |
| 1989-93 | 989 | 1,319 |
| 1990-94 | 936 | 1,232 |
| 1991-95 | 863 | 1,106 |
| 1992-96 | 779 | 950 |
| 1993-97 | 794 | 923 |
| 1994-98 | 866 | 962 |
| 1995-99 | 1,043 | 1,110 |
| 1996-00 | 1,246 | 1,271 |
| 1997-01 | 1,428 | 1,442 |
| 1998-02 | 1,568 | 1,568 |
| 1999-03 | 1,660 | 1,660 |
| 2000-04 | 1,662 | 1,629 |
| 2001-05 | 1,568 | 1,493 |
| 2002-06 | 1,449 | 1,329 |
| 2003-07 | 1,342 | 1,167 |
| 2004-08 | 1,283 | 1.043 |

TABLE S4. Number of all publications from *s*=1 to *s*=20, with *cs*(max)=1, *ca*(min)=5, *a*(min)=*a*(max)=5, within period 1980-2008.

| s | N | last publ y | #publ y |
|---|---|---|---|
| 1 | 501,671 | 2008 | 29 |
| 2 | 170,672 | 2008 | 29 |
| 3 | 39,903 | 2008 | 29 |
| 4 | 12,650 | 2008 | 29 |
| 5 | 5,508 | 2008 | 29 |
| 6 | 2,887 | 2007 | 28 |
| 7 | 1,718 | 2006 | 27 |
| 8 | 1,178 | 2005 | 26 |
| 9 | 858 | 2004 | 25 |
| 10 | 639 | 2003 | 24 |
| 11 | 505 | 2002 | 23 |
| 12 | 412 | 2001 | 22 |
| 13 | 345 | 2000 | 21 |
| 14 | 292 | 1999 | 20 |
| 15 | 215 | 1998 | 19 |
| 16 | 199 | 1997 | 18 |
| 17 | 178 | 1996 | 17 |
| 18 | 164 | 1995 | 16 |
| 19 | 124 | 1994 | 15 |
| 20 | 120 | 1993 | 14 |

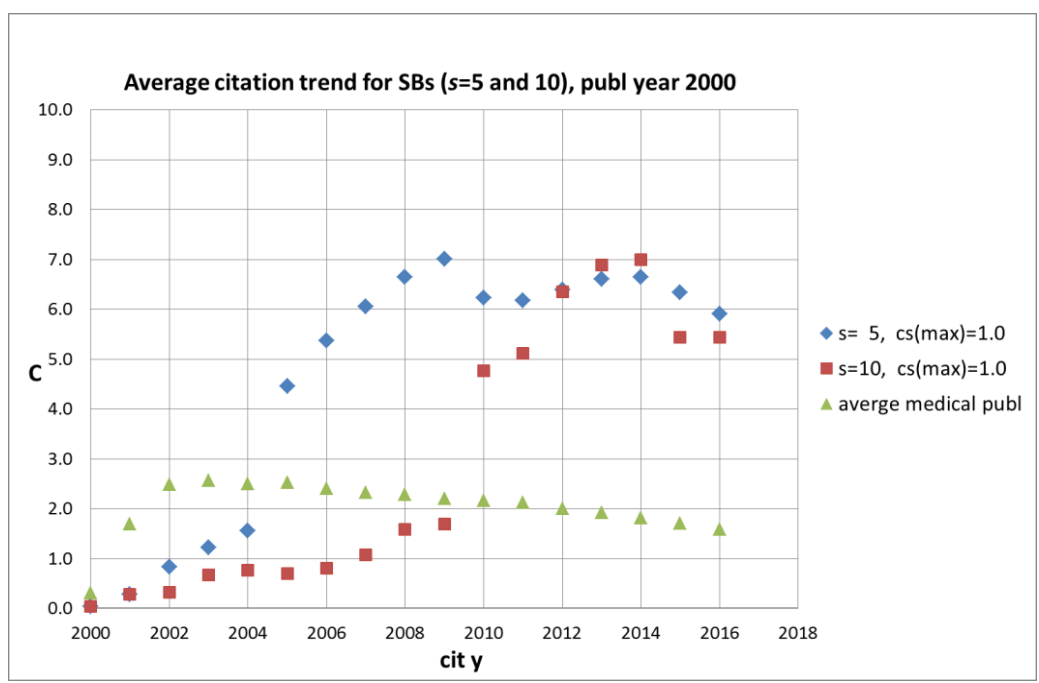

FIG. S1. Average citation trend in the period 2000-2016 for SBs with *s*=5 and 10 (both *cs*(max)=1.0) as well an average medical publication with publication year 2000.

TABLE S5. Number of SBs (*s*=5) for successive during-sleep citation-intensity intervals

| *s=5* | *cs* | | | | | |
|---|---|---|---|---|---|---|
| *pub y* | **0.0** | **0.2** | **0.4** | **0.6** | **0.8** | **1.0** |
| 1980-84 | 5 | 11 | 24 | 48 | 75 | 128 |
| 1981-85 | 4 | 11 | 29 | 57 | 72 | 140 |
| 1982-86 | 3 | 15 | 32 | 63 | 77 | 154 |
| 1983-87 | 4 | 17 | 36 | 83 | 93 | 177 |
| 1984-88 | 3 | 19 | 45 | 98 | 111 | 201 |
| 1985-89 | 2 | 27 | 59 | 110 | 150 | 243 |
| 1986-90 | 6 | 35 | 71 | 125 | 201 | 293 |
| 1987-91 | 11 | 43 | 89 | 158 | 247 | 348 |
| 1988-92 | 17 | 48 | 95 | 162 | 259 | 389 |
| 1989-93 | 19 | 49 | 94 | 163 | 269 | 395 |
| 1990-94 | 19 | 45 | 86 | 168 | 244 | 374 |
| 1991-95 | 16 | 37 | 77 | 164 | 225 | 344 |
| 1992-96 | 14 | 32 | 62 | 143 | 201 | 327 |
| 1993-97 | 8 | 33 | 65 | 140 | 205 | 343 |
| 1994-98 | 8 | 39 | 74 | 154 | 212 | 379 |
| 1995-99 | 10 | 46 | 96 | 173 | 272 | 446 |
| 1996-00 | 13 | 54 | 110 | 201 | 343 | 525 |
| 1997-01 | 12 | 54 | 120 | 231 | 405 | 606 |
| 1998-02 | 14 | 53 | 135 | 258 | 444 | 664 |
| 1999-03 | 14 | 45 | 133 | 270 | 484 | 714 |
| 2000-04 | 13 | 41 | 123 | 265 | 488 | 732 |
| 2001-05 | 13 | 39 | 114 | 233 | 436 | 733 |
| 2002-06 | 12 | 34 | 108 | 216 | 395 | 684 |
| 2003-07 | 14 | 32 | 89 | 199 | 366 | 642 |

TABLE S6.  Number of SBs (***s***=10) for successive during-sleep citation-intensity intervals

|  |  | ***cs*** |  |  |  |  |  |  |  |  |  |
|---|---|---|---|---|---|---|---|---|---|---|---|
|  | ***s***=10 | *0* | *0.2* | *0.3* | *0.4* | *0.5* | *0.6* | *0.7* | *0.8* | *0.9* | *1.0* |
| 1980-84 | 1982 | 0 | 1 | 4 | 3 | 4 | 7 | 7 | 9 | 13 | 27 |
| 1981-85 | 1983 | 0 | 4 | 5 | 6 | 7 | 7 | 11 | 10 | 20 | 30 |
| 1982-86 | 1984 | 0 | 4 | 6 | 11 | 10 | 11 | 16 | 15 | 21 | 35 |
| 1983-87 | 1985 | 0 | 4 | 7 | 12 | 12 | 9 | 18 | 18 | 26 | 38 |
| 1984-88 | 1986 | 0 | 3 | 5 | 10 | 12 | 12 | 21 | 17 | 31 | 44 |
| 1985-89 | 1987 | 0 | 4 | 3 | 9 | 12 | 11 | 18 | 15 | 33 | 38 |
| 1986-90 | 1988 | 0 | 2 | 3 | 6 | 8 | 14 | 15 | 14 | 28 | 33 |
| 1987-91 | 1989 | 0 | 2 | 4 | 2 | 7 | 11 | 14 | 8 | 27 | 27 |
| 1988-92 | 1990 | 0 | 2 | 4 | 3 | 4 | 13 | 14 | 9 | 24 | 27 |
| 1989-93 | 1991 | 1 | 2 | 4 | 4 | 3 | 10 | 12 | 9 | 19 | 23 |
| 1990-94 | 1992 | 1 | 1 | 5 | 5 | 3 | 11 | 11 | 12 | 20 | 25 |
| 1991-95 | 1993 | 1 | 0 | 5 | 7 | 3 | 8 | 15 | 14 | 25 | 29 |
| 1992-96 | 1994 | 1 | 0 | 3 | 7 | 1 | 8 | 15 | 20 | 34 | 35 |
| 1993-97 | 1995 | 2 | 0 | 3 | 6 | 3 | 12 | 21 | 20 | 42 | 35 |
| 1994-98 | 1996 | 3 | 0 | 4 | 5 | 5 | 12 | 27 | 28 | 50 | 44 |
| 1995-99 | 1997 | 3 | 0 | 5 | 7 | 8 | 13 | 29 | 33 | 58 | 46 |
| 1996-00 | 1998 | 3 | 1 | 5 | 7 | 9 | 12 | 29 | 36 | 61 | 51 |
| 1997-01 | 1999 | 4 | 1 | 5 | 7 | 9 | 13 | 31 | 34 | 56 | 52 |
| 1998-02 | 2000 | 4 | 3 | 4 | 7 | 8 | 9 | 27 | 33 | 48 | 51 |

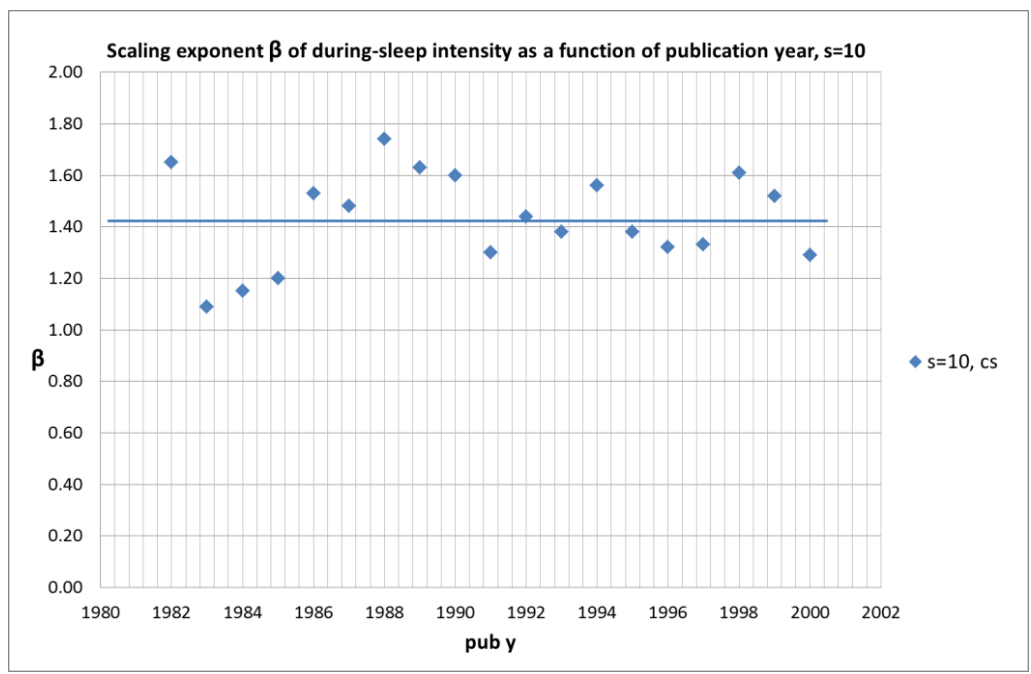

FIG. S2. Scaling exponent **β** of during-sleep citation intensity (SB-SNPRs with ***s***=10), each five-years block is located in the figure by its middle year.

TABLE S7. Number of SBs (*s*=5) for awake citation-intensity intervals

|  | *ca* *s*=5 | 5.0≤ *ca* ≤6.0 | 6.0< *ca* ≤7.0 | 7.0< *ca* ≤9.0 | 9.0< *ca* ≤11.0 | 11.0< *ca* ≤13.0 | 13.0< *ca* ≤15.0 | 15.0< *ca* ≤17.0 |
|---|---|---|---|---|---|---|---|---|
| 1980-84 | 1982 | 191 | 40 | 41 | 11 | 4 | 1 | 2 |
| 1981-85 | 1983 | 203 | 47 | 41 | 13 | 4 | 1 | 2 |
| 1982-86 | 1984 | 220 | 51 | 50 | 13 | 5 | 0 | 2 |
| 1983-87 | 1985 | 272 | 53 | 59 | 17 | 4 | 1 | 1 |
| 1984-88 | 1986 | 331 | 64 | 56 | 17 | 3 | 3 | 0 |
| 1985-89 | 1987 | 397 | 97 | 69 | 16 | 5 | 3 | 2 |
| 1986-90 | 1988 | 480 | 126 | 93 | 19 | 4 | 4 | 2 |
| 1987-91 | 1989 | 576 | 167 | 113 | 25 | 6 | 5 | 3 |
| 1988-92 | 1990 | 622 | 189 | 122 | 25 | 6 | 4 | 4 |
| 1989-93 | 1991 | 637 | 192 | 123 | 25 | 8 | 2 | 4 |
| 1990-94 | 1992 | 611 | 176 | 115 | 24 | 6 | 3 | 2 |
| 1991-95 | 1993 | 566 | 166 | 100 | 20 | 6 | 3 | 2 |
| 1992-96 | 1994 | 528 | 144 | 85 | 15 | 4 | 2 | 1 |
| 1993-97 | 1995 | 540 | 152 | 85 | 9 | 3 | 2 | 0 |
| 1994-98 | 1996 | 585 | 170 | 98 | 7 | 1 | 3 | 0 |
| 1995-99 | 1997 | 718 | 200 | 108 | 11 | 3 | 2 | 1 |
| 1996-00 | 1998 | 880 | 226 | 120 | 15 | 5 | 2 | 1 |
| 1997-01 | 1999 | 997 | 264 | 141 | 22 | 5 | 3 | 1 |
| 1998-02 | 2000 | 1088 | 293 | 149 | 28 | 7 | 6 | 2 |
| 1999-03 | 2001 | 1157 | 307 | 149 | 34 | 9 | 8 | 3 |
| 2000-04 | 2002 | 1146 | 304 | 167 | 33 | 8 | 8 | 2 |
| 2001-05 | 2003 | 1070 | 297 | 159 | 28 | 8 | 9 | 2 |
| 2002-06 | 2004 | 1007 | 267 | 139 | 24 | 7 | 8 | 2 |
| 2003-07 | 2005 | 940 | 248 | 127 | 23 | 5 | 6 | 1 |

TABLE S8. Number of SBs (*s*=10) for awake citation-intensity intervals

|  | *ca* |  |  |  |  |  |
|---|---|---|---|---|---|---|
|  | s=10 | 5.0≤ *ca* ≤6.0 | 6.0< *ca* ≤7.0 | 7.0< *ca* ≤9.0 | 9.0< *ca* ≤11.0 | 11.0< *ca* ≤15.0 |
| 1980-84 | 1982 | 42 | 22 | 9 | 1 | 1 |
| 1981-85 | 1983 | 56 | 31 | 12 | 0 | 1 |
| 1982-86 | 1984 | 77 | 35 | 14 | 2 | 1 |
| 1983-87 | 1985 | 88 | 37 | 15 | 2 | 2 |
| 1984-88 | 1986 | 94 | 41 | 15 | 3 | 2 |
| 1985-89 | 1987 | 86 | 38 | 11 | 5 | 3 |
| 1986-90 | 1988 | 77 | 27 | 10 | 5 | 4 |
| 1987-91 | 1989 | 63 | 25 | 7 | 3 | 4 |
| 1988-92 | 1990 | 64 | 22 | 7 | 3 | 4 |
| 1989-93 | 1991 | 58 | 17 | 7 | 2 | 3 |
| 1990-94 | 1992 | 66 | 17 | 9 | 0 | 2 |
| 1991-95 | 1993 | 73 | 21 | 11 | 1 | 1 |
| 1992-96 | 1994 | 86 | 20 | 14 | 2 | 2 |
| 1993-97 | 1995 | 100 | 25 | 15 | 3 | 1 |
| 1994-98 | 1996 | 122 | 29 | 21 | 6 | 2 |
| 1995-99 | 1997 | 136 | 31 | 26 | 6 | 3 |
| 1996-00 | 1998 | 146 | 33 | 25 | 7 | 3 |
| 1997-01 | 1999 | 145 | 33 | 22 | 8 | 4 |
| 1998-02 | 2000 | 133 | 28 | 22 | 7 | 4 |

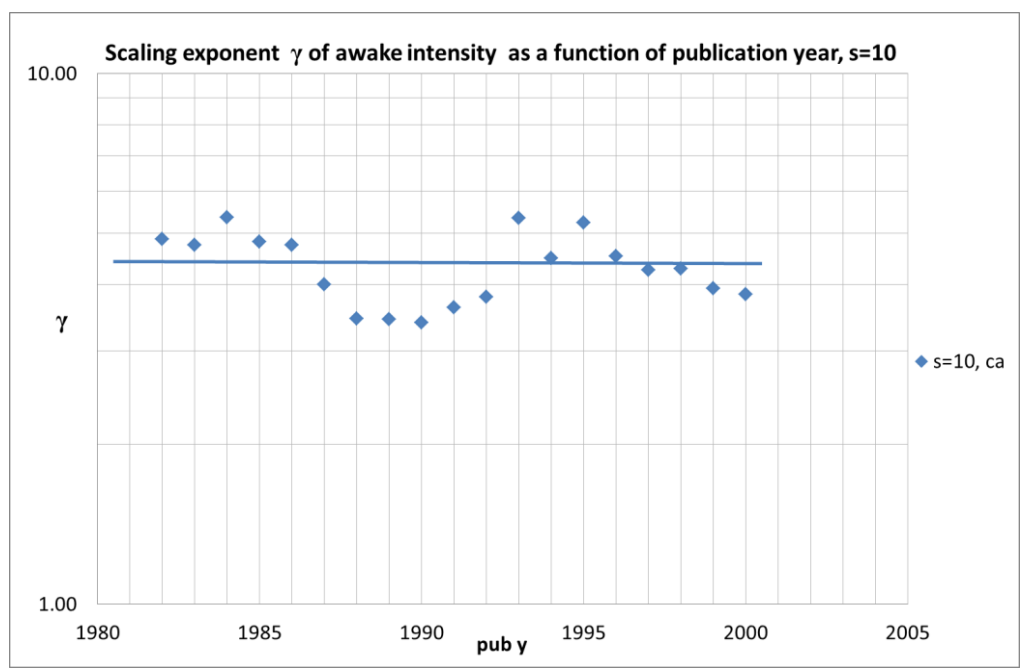

FIG. S3. Scaling exponent γ of awake citation-intensity (SB-SNPRs with *s*=10), each five-years block is located in the figure by its middle year.